\documentclass[aps,prd,amsmath,floats,floatfix, twocolumn,
superscriptaddress,nofootinbib,showpacs]{revtex4-1}

\usepackage[colorlinks, pdfborder={0 0 0}, plainpages=false]{hyperref}
\usepackage{graphicx}
\usepackage{xspace}
\usepackage[usenames,dvipsnames]{color}
\usepackage{amssymb, amsmath}
\usepackage[normalem]{ulem} 
\usepackage[caption=false]{subfig}
\usepackage{multirow}

\usepackage{blindtext}

\DeclareMathAlphabet{\mathpzc}{OT1}{pzc}{m}{it}

\newcommand{\ba}{\overline}

\newcommand{\h}{H}
\newcommand{\hmath}{\mathpzc{h}}

\newcommand{\Lb}{\mathbb{L}}

\usepackage{xspace}
\newcommand{\SKSHorBC}{SKS-Eq-$\theta_0$\xspace}
\newcommand{\SKS}{SKS-Eq\xspace}
\newcommand{\SHK}{SH-H\xspace}
\newcommand{\SDH}{SDH-DH\xspace}
\newcommand{\SKSDH}{SKS-DH\xspace}

\newcommand{\Caltech}{\affiliation{Theoretical Astrophysics 350-17,
    California Institute of Technology, Pasadena, CA 91125, USA}}

\newcommand{\CITA}{\affiliation{Canadian Institute for Theoretical
    Astrophysics, 60 St.~George Street, University of Toronto,
    Toronto, ON M5S 3H8, Canada}} %
\newcommand{\AEI}{\affiliation{Albert-Einstein-Institut,
	Max-Planck-Institut für Gravitationsphysik,
D-14476 Potsdam-Golm, Germany}} %

\begin{document}
\title{
Comparison of binary black hole initial data sets
}

\author{Vijay Varma} \Caltech
\author{Mark A. Scheel} \Caltech
\author{Harald P. Pfeiffer} \AEI \CITA

\date{\today}

\begin{abstract}
We present improvements to construction of binary black hole initial data
used in SpEC (the Spectral Einstein Code). We introduce new boundary
conditions for the extended conformal thin sandwich elliptic equations that
enforce the excision surfaces to be slightly inside rather than on the
apparent horizons, thus avoiding extrapolation into the black holes at the
last stage of initial data construction. We find that this improves initial
data constraint violations near and inside the apparent horizons by about 3
orders of magnitude. We construct several initial data sets that are intended
to be astrophysically equivalent but use different free data, boundary
conditions, and initial gauge conditions. These include free data chosen as a
superposition of two black holes in time-independent horizon-penetrating
harmonic and damped harmonic coordinates. We also implement initial data for
which the initial gauge satisfies the harmonic and damped harmonic gauge
conditions; this can be done independently of the free data, since this
amounts to a choice of the time derivatives of the lapse and shift. We compare
these initial data sets by evolving them. We show that the gravitational
waveforms extracted during the evolution of these different initial data sets
agree very well after excluding initial transients. However, we do find small
differences between these waveforms, which we attribute to small differences
in initial orbital eccentricity, and in initial BH masses and spins, resulting
from the different choices of free data. Among the cases considered, we find
that superposed harmonic initial data leads to significantly smaller
transients, smaller variation in BH spins and masses during these transients,
smaller constraint violations, and more computationally efficient evolutions.
Finally, we study the impact of initial data choices on the construction of
zero-eccentricity initial data.
\end{abstract}


\maketitle

\section{Introduction}
Numerical simulations of binary black holes (BBH) have been crucial for our
understanding of BBH systems. For example, these simulations are important for
the construction of accurate waveform models that cover the
inspiral-merger-ringdown phases of a BBH system~\cite{Khan:2015jqa,
Hannam:2013oca, Bohe:2016gbl, Pan:2013rra, Taracchini:2013rva}; these
models were used in successful detections~\cite{LIGOVirgo2016a,
Abbott:2016nmj, Abbott:2017gyy, Abbott:2017oio, Abbott:2017vtc} of
gravitational waves by LIGO~\cite{aLIGO2}. Accurate waveform models are
necessary not only for the detection of gravitational wave signals but also
for making inferences about the astrophysical properties of the
sources~\cite{TheLIGOScientific:2016wfe} and for conducting strong field tests
of general relativity~\cite{TheLIGOScientific:2016src}.

A numerical BBH simulation begins with the construction of initial data that
describes the state of the system on some three-dimensional initial surface
labeled $t=0$. Constructing initial data requires not only solving the
Einstein constraint equations, but also freely choosing the initial spatial
coordinates, the embedding of the three-dimensional initial surface in the
four-dimensional spacetime, and some physical degrees of freedom; these
choices are encoded in freely-specifiable functions and boundary conditions
that are used in the solution of the constraint equations.  The subset of
these choices that amount to choosing coordinates should not, of course,
affect the physics~\cite{Garcia:2012dc}, but they may affect the robustness
and accuracy of the subsequent evolution. This is because they influence the
gauge degrees of freedom that evolve along with, and are intermixed with, the
physical degrees of freedom.

In this paper we study how binary black hole simulations are affected by
different choices of free data, gauge, and boundary conditions that
are made when constructing initial data sets that are meant to be physically
identical. We consider simulations performed with one particular numerical
relativity code, the Spectral Einstein Code (SpEC) \cite{SpECWebsite}.

\subsection{Summary of initial data for SpEC simulations}
Before discussing how to improve the treatment of initial data, we first
outline the current procedure used to construct initial data for binary black
hole simulations using SpEC; this procedure is described in more detail
in Sec.~\ref{Sec:IDFormalism}. We adopt the Extended Conformal
Thin Sandwich (XCTS) formalism ~\cite{York1999, Pfeiffer2003b}, and the free
data supplied to the XCTS equations are chosen to be constructed from a
superposition of two single black holes (BHs) in Kerr-Schild
coordinates~\cite{Lovelace2008}. The region inside each of the BHs is excised
from the computational domain, and boundary conditions are chosen that enforce
the boundaries of these excision regions to be apparent
horizons~\cite{Cook2004}.

After the XCTS system of equations is solved, yielding a constraint-satisfying
initial data set, the metric quantities are interpolated (and extrapolated)
onto a new numerical grid that extends slightly inside the original excision
boundaries. This new grid is used for the evolution. On the new grid the
apparent horizons lie inside the computational domain
rather than on its boundary, and this allows the
subsequent evolution to track the apparent horizons
as they dynamically change in shape and size.
Unfortunately, the small extrapolation to points inside the
apparent horizons introduces some constraint violations in the vicinity of
the excision boundaries.

Binary black hole initial data described above represent a
physical solution to Einstein's equations but do not result in an
exact snapshot of a quasi-equilibrium inspiral: the solution
contains near-zone transient dynamics and does not include the
correct initial gravitational radiation in the far zone.  During
evolution the system relaxes into a quasi-equilibrium state with the
mismatch radiating away as a pulse of spurious radiation, which is
generally referred to as \emph{junk radiation}.  The initial transients
typically contain high spatial and temporal frequencies,
so that resolving them is computationally expensive.  For this
reason, we typically choose not to fully resolve them at all, and we
instead simply discard the initial part of the gravitational
waveforms that are affected by these transients.

In addition to initial data, evolution also requires an initial choice of
gauge. SpEC employs the generalized harmonic formulation of the Einstein
equations~\cite{Lindblom:2007, Friedrich1985, Garfinkle2002, Pretorius2005c},
where gauge conditions are imposed through gauge source functions $H_a$ (see
\ref{Subsec:IDGaugeChoices}). At the beginning of a binary black hole
simulation, $H_a$ is currently chosen such that the time derivatives of lapse
and shift vanish at $t=0$ in a frame co-rotating with the binary; this
quasi-equilibrium condition is intended to minimize gauge dynamics at the
beginning of the evolution~\cite{Scheel2009}.  However, a different choice of
$H_a$, the {\it damped harmonic} gauge~\cite{Lindblom2009c,Choptuik:2009ww,
Szilagyi:2009qz}, is usually necessary later in the evolution when the black
holes merge. The choice of $H_a$ cannot be discontinuous in time because time
derivatives of $H_a$ appear in the evolution equations.  Hence, a smooth gauge
transformation is applied in the early stages of evolution to move into damped
harmonic gauge.

\subsection{Improvements in initial data treatment}

In this paper we present several improvements to BBH initial data
construction. First, we introduce new boundary conditions for the XCTS
elliptic equations that enforce the excision surfaces to have a negative
expansion. This means that the excision surfaces are already inside
the apparent horizons, eliminating the
need to extrapolate inside the horizons during
the initial data construction. We find that this improves
constraint violations in initial data near and inside the apparent horizon
surfaces by about 3 orders of magnitude.

Next, we construct several initial data sets that
implement different free data in the XCTS equations as well as different
initial gauge conditions. The new free data choices include
superpositions of two single BHs in time-independent horizon-penetrating
harmonic~\cite{cook_scheel97} and damped harmonic~\cite{Varma:2018sbhdh}
coordinates rather than in Kerr-Schild coordinates.
The new initial gauge choices include imposing (to numerical truncation
error) the harmonic and damped harmonic
gauge conditions at $t=0$, instead of setting the initial
time derivatives of the lapse and shift to zero.

We evolve all these initial data sets.
Among all the initial data constructions considered here, we find that
superposed harmonic initial data exhibits the most favorable behavior
in subsequent evolutions.
Superposed harmonic initial data exhibits
the smallest amount of junk radiation,
and the smallest variation in the measured masses and spins of the BHs
during the initial relaxation. Furthermore, the constraint violations during the initial
relaxation are smaller
by about an order of magnitude. Remarkably, evolution of superposed harmonic
initial data also shows a speed-up of about $33\%$ compared to
superposed Kerr-Schild data for the case considered, reducing the runtime and
computational cost of BBH simulations. The speed-up can be traced to the
adaptive mesh refinement (AMR) choosing fewer grid points
to achieve the same accuracy.
We also find that during the initial relaxation, when we intentionally do
not attempt to resolve initial transients,
the constraint violations converge to zero with increasing
resolution only for superposed harmonic initial data.

These positive findings
suggest that simulations in the future
should use superposed harmonic initial data; however, it is known that a
single BH in time-independent horizon-penetrating harmonic coordinates
becomes very distorted in the direction of spin
for large spins (cf. Fig.~\ref{Fig:Pancake}).
These distortions are inherited by the superposed harmonic BBH initial data
sets, so that the black hole horizons become so deformed to render evolutions
of nearly extremal spins impractical. We find that superposed harmonic initial
data works well when both BH dimensionless spin magnitudes are below $0.7$.

We also find that superposed damped harmonic initial data does not
perform as well as superposed Kerr-Schild initial data in the above
respects.  However, we find that we can construct superposed
Kerr-Schild initial data that is initially in damped harmonic gauge
(so as to avoid a subsequent gauge transformation during the
evolution), and that this initial data set performs as well as
superposed Kerr-Schild with the current quasiequilbrium initial gauge,
in the above respects.
Therefore, we recommend that superposed harmonic initial data be used
for spin magnitudes $\leq 0.7$.  For higher spins, we recommend
superposed Kerr-Schild initial data with damped harmonic initial gauge,
since this performs no
worse than the current choice of superposed Kerr-Schild with
quasi-equilibrium initial gauge, and it is simpler because it requires no
gauge transition during evolution.

The rest of the paper is organized as
follows. Section~\ref{Sec:IDFormalism} provides a brief overview of
the initial data formalism, including the new negative-expansion
boundary conditions and new choices of free data and initial gauge. In
Sec.~\ref{Sec:IDTypes} we summarize the particular choices of initial
data that we choose to construct and compare in this work.  In
Sec.~\ref{Sec:IDTests} we test convergence of constraints in each of
these initial data sets.
In Sec.~\ref{Sec:BBhEv} we evolve these different initial data sets
and compare the results of these evolutions. Finally, in
Sec.~\ref{Sec:Conclusion} we provide a conclusion and recommendations
for the construction of initial data in future BBH
simulations. Throughout this paper we use geometric units with
$G=c=1$. We use Latin letters from the start of the alphabet
$(a,b,c,\dots)$ for spacetime indices and from the middle of the
alphabet $(i,j,k,\dots)$ for spatial indices.  We use $\psi_{ab}$ for
the space-time metric. We use $g_{ij}$ for the spatial metric, $N$ for
the lapse and $N^i$ for the shift of the constant-$t$ hypersurfaces.

We note that this paper focuses entirely on improvements to the initial
data treatment adopted by codes~\cite{SpECWebsite, Pretorius2005c} that use
the generalized harmonic formulation~\cite{Lindblom:2007, Friedrich1985,
  Garfinkle2002, Pretorius2005c} of the evolution equations.  NR
codes~\cite{Baumgarte99, Bruegmann2006, Zlochower:2005bj, Sperhake2006,
  Pollney:2009yz, Herrmann2007b, Pekowsky:2013ska} that use moving-puncture
initial data~\cite{Brandt1997} (since they do not employ BH excision)
and/or the BSSNOK formulation~\cite{Baumgarte99, shibata95, NOK87} of the
evolution equations (since the gauge is set directly by setting a lapse and
a shift, rather than a gauge souce funcion) would not benefit from these
improvements.

\section{BBH Initial Data Formalism}
\label{Sec:IDFormalism}
In this section we provide a brief overview of binary black hole initial data
formalism, and we suggest improved boundary conditions and gauge choices. We
start by discussing the Extended Conformal Thin Sandwich (XCTS) system of
elliptic equations in Sec.~\ref{Subsec:XCTS}. Next, in
Sec.~\ref{Subsec:IDBdryCond} we cover the boundary conditions for the elliptic
equations, including the new negative expansion boundary conditions that lets
us avoid spatial extrapolation of the initial data quantities. Finally, in
Sec.~\ref{Subsec:IDGaugeChoices} we discuss different gauge choices that we
use in initial data. In the next section, Sec.~\ref{Sec:IDTypes}, we summarize
the different initial data sets constructed for this study.

\subsection{Extended conformal thin sandwich equations}
\label{Subsec:XCTS}

XCTS~\cite{York1999, Pfeiffer2003b}
is a formulation of the Einstein constraint equations
well-suited for numerical solution. The ``extended'' part of XCTS refers to
an additional equation that is added to the system: the evolution equation
for the trace of the extrinsic curvature, converted into an elliptic equation.
This extra equation is useful in producing initial data in quasi-equilibrium.
For a more detailed review of initial data construction,
see \cite{baumgarteShapiroBook,  Pfeiffer:2005, Cook2000}.

The XCTS construction starts with a conformal decomposition
of the 3-metric into a conformal factor $\psi$ and a conformal metric
$\bar{g}_{ij}$
\begin{gather}
g_{ij} = \psi^4~\bar{g}_{ij}.
\label{Eq:Conformalg}
\end{gather}

Using the definition of extrinsic curvature in terms of the time derivative
of the spatial metric, the extrinsic curvature $K_{ij}$ takes the form
\begin{gather}
K_{ij} = \frac{1}{3} g_{ij} K + A_{ij},
\end{gather}
where
\begin{gather}
A_{ij} = \psi^{-2} \bar{A}_{ij}, ~~~
\bar{A}^{ij} = \frac{\psi^6}{2 N} \left( (\bar{\Lb} N)^{ij}
- \bar{u}^{ij} \right).
\label{Eq:ConformalK}
\end{gather}
Here $N$ is the lapse, $N^i$ is the shift, $(\bar{\Lb} N)^{ij}$ represents the
conformal Killing operator in conformal space, and
$\bar{u}_{ij} = \partial_t \bar{g}_{ij}$\footnote{Note that one also needs to
set $\bar{g}^{ij} \bar{u}_{ij}=0$ to uniquely specify $\ba{u}_{ij}$.}.
$K$ and $A_{ij}$ are the trace and trace-free part of $K_{ij}$.

In the XCTS formalism, one can freely specify the conformal
metric $\bar{g}_{ij}$, trace of extrinsic curvature
$K$, and
their time derivatives $\bar{u}_{ij}$ and $\partial_t K$. For
quasi-equilibrium situations, these time derivatives are typically set to zero.
The system of of elliptic equations to be solved becomes:
\begin{gather}
\label{Eq:xctsConf}
\bar{\nabla}^2 \psi - \frac{1}{8} \bar{R} \psi - \frac{1}{12} K^2 \psi^5
+\frac{1}{8} \psi^{-7} \bar{A}^{ij} \bar{A}_{ij} = 0, \\
\label{Eq:xctsShift}
\bar{\nabla}_j \left(\frac{\psi^6}{2 N} (\bar{\Lb} N)^{ij}\right)
-\frac{2}{3} \psi^6 \bar{\nabla}^i K
-\bar{\nabla}_j \left(\frac{\psi^6}{2 N} \bar{u}^{ij} \right) = 0, \\
\bar{\nabla}^2 (N \psi)
-N \psi \left(\frac{\bar{R}}{8} +\frac{5}{12} K^4 \psi^4
+\frac{7}{8} \psi^{-8} \bar{A}^{ij} \bar{A}_{ij} \right) \nonumber \\
+\psi^5 (\partial_t K - N^k \partial_k K) = 0,
\label{Eq:xctsLapse}
\end{gather}
where $\bar{R}$ and $\bar{\nabla}_i$ are
the Ricci scalar and the spatial covariant derivative operator associated
with $\bar{g}_{ij}$. Once these equations are solved for $\psi$, $N \psi$
and $N^i$, the physical solution ($g_{ij}$, $K_{ij}$) is constructed from
Eqs.~(\ref{Eq:Conformalg}-\ref{Eq:ConformalK}) and the free data
($\bar{g}_{ij}$, $\bar{u}_{ij}$, $K$ and $\partial_t K$).

\subsubsection{Choosing freely specifiable data}
\label{Subsubsec:XCTSFreeData}

If the lapse $N$ and shift $N^i$ computed from XCTS are used in the evolution
of the initial data, the time derivative of $K$ will initially be equal to the
specified $\partial_t K$ and the trace-free part of $\partial_t g_{ij}$ will
be initially proportional to the specified $\bar{u}_{ij}$. In order to
generate quasi-equilibrium initial data, the natural choice for these freely
specifiable quantities is:
\begin{gather}
\label{Eq:dtgID}
\bar{u}_{ij} = 0, ~~~~ \partial_t K = 0.
\end{gather}

Following Ref.~\cite{Lovelace2008},
we construct the free data based on a superposition of two single-BH
solutions. Let $g^{\alpha}_{ij}$ and $K^{\alpha}$ be the 3-metric and
the trace of extrinsic curvature of a single boosted, spinning black
hole, with $\alpha = 1,2$ labeling the two black holes.  We then
choose the conformal 3-metric $\bar{g}_{ij}$
and the trace of the extrinsic curvature $K$ to be
\begin{gather}
\label{Eq:Superposition}
\ba{g}_{ij} = f_{ij} +\sum_{\alpha=1}^2 e^{-r_{\alpha}^2/w_{\alpha}^2}
~(g^{\alpha}_{ij} -f_{ij}), \\
K = \sum_{\alpha=1}^2 e^{-r_{\alpha}^2/w_{\alpha}^2} K^{\alpha},
\label{Eq:SuperpositionTrK}
\end{gather}
where $f_{ij}$ is the flat 3-metric. Far from the holes, the conformal metric
is very nearly flat and the trace of extrinsic curvature is very nearly zero.
This is achieved through a Gaussian weight around each hole,
with a width $w_{\alpha}$ that determines how fast the conformal metric
approaches the flat metric with increasing Euclidean distance $r_{\alpha}$
from the center of each hole. The widths of the Gaussians $w_{\alpha}$ are
chosen to be
\begin{gather}
w_{\alpha} = 0.6~ d^{L_1}_{\alpha},
\end{gather}
where $d^{L_1}_{\alpha}$ is the Euclidean distance to
the Newtonian $L_1$ Lagrange point from the center of hole $\alpha$.
This is identical to the choice made in Ref.~\cite{Lovelace2008}.
This ensures that the widths are larger than the size scale of the hole
($\sim M_{\alpha}$, the mass of the hole) but smaller than the distance to
the other hole. This also ensures that near each black hole, the contributions
of the other black hole are attenuated by several orders of magnitude. The
Gaussians are also needed so that at large distances the solution does not
develop a logarithmic singularity~\cite{Lovelace2009}.

The single-BH quantities $g^{\alpha}_{ij}$ and $K^{\alpha}$
above are determined by the Kerr metric, by a choice of how to slice
the Kerr metric into a foliation of three-dimensional hypersurfaces,
and by a choice of spatial coordinates on these hypersurfaces.
These choices are largely arbitrary, but they must satisfy certain
conditions to produce a viable initial data set; for example, the
slices must contain an apparent horizon and be regular there.

\subsubsection{Exploring new choices of free data}
A key goal of this paper is to investigate the effect of the choice
of $g^{\alpha}_{ij}$ and $K^{\alpha}$
on the resulting initial data set and subsequent evolution.  Here we
consider three choices, explained in more detail in Sec.~\ref{Sec:IDTypes}.
The first is the choice made in
the current implementation of SpEC, which was introduced in
Ref.~\cite{Lovelace2008}: $g^{\alpha}_{ij}$ and
$K^{\alpha}$ are taken to be in Kerr-Schild coordinates centered about each
BH. The second is to specify
$g^{\alpha}_{ij}$ and $K^{\alpha}$ in
harmonic coordinates, using the unique harmonic
time slicing that is both time-independent (for a single BH) and that
penetrates the horizon as derived in Ref.~\cite{cook_scheel97}.
Finally, we also consider the case in
which $g^{\alpha}_{ij}$ and $K^{\alpha}$ are chosen in the unique
coordinate system that obeys the damped harmonic
condition~\cite{Lindblom2009c, Choptuik:2009ww, Szilagyi:2009qz} and
for which the time slices are time-independent
and horizon-penetrating~\cite{Varma:2018sbhdh}.
For all of these cases, we use the same
Gaussian weights in Eqs.~(\ref{Eq:Superposition})
and~(\ref{Eq:SuperpositionTrK}).

\subsection{Boundary conditions}
\label{Subsec:IDBdryCond}

Equations~(\ref{Eq:xctsConf}), (\ref{Eq:xctsShift}),
and~(\ref{Eq:xctsLapse}) require appropriate boundary conditions in order to
solve for initial data.

The outer boundary (denoted by $\mathcal{B}_{\infty}$) conditions are obtained
by requiring the initial data to be asymptotically flat. Note that in
practice, we do not actually place the boundary $\mathcal{B}_{\infty}$ at
spatial infinity, but at a coordinate sphere of radius $\sim 10^9 M$.
Because the conformal metric and trace of extrinsic curvature,
as given by Eqs.~(\ref{Eq:Superposition}) and~(\ref{Eq:SuperpositionTrK}),
are already asymptotically flat, the outer boundary conditions are
\begin{gather}
\psi = 1 ~~~~~\text{at $\mathcal{B}_{\infty}$}, \\
N \psi = 1 ~~~\text{at $\mathcal{B}_{\infty}$}, \\
\widetilde{N}^i = (\bold{\Omega_0} \times \bold{r})^i + \dot{a}_0 r^i
~~~\text{at $\mathcal{B}_{\infty}$}.
\label{Eq:ShiftOuterBC}
\end{gather}
Here, $\widetilde{N}^i$ is the shift in a frame that co-rotates with the
binary, $r^i$ is the coordinate position vector, $\bold{\Omega_0}$ is the
orbital angular velocity and $\dot{a}_0$ is an expansion parameter.
The shift boundary condition consists of a rotation and an expansion term.
The rotation term (parametrized by $\bold{\Omega_0}$) ensures that
the time coordinate is helical and tracks the rotation of the system, and
the expansion term (parametrized by $\dot{a}$) sets a non-zero
radial velocity, to account for the initial decrease in the orbit due to
radiation reaction.
These boundary conditions are identical to those
in~\cite{Pfeiffer-Brown-etal:2007},
which presents a more detailed exposition.

The inner boundary conditions are imposed on the excision surfaces,
denoted by $\mathcal{B}_{E}$.  These are chosen to be
surfaces of constant
radial coordinate in the single BH coordinates used in
Eq.~\ref{Eq:Superposition}.  We choose our single BH coordinates such that
the apparent horizon has a constant radial coordinate\footnote{For superposed
Kerr-Schild and superposed harmonic, this is the Boyer-Lindquist radius;
for superposed damped harmonic, this coordinate is determined
numerically~\cite{Varma:2018sbhdh}.} but the excision boundary may or may not be
an apparent horizon, as explained below. Here we consider two types of
inner boundary conditions.

\subsubsection{Horizon boundary conditions}
\label{Subsubsec:HorBC}

The standard practice in SpEC has been to choose quasi-equilibrium
apparent/isolated horizon boundary conditions on the inner excision
surfaces\cite{Cook2002,Cook2004}.
We refer the reader to \cite{Ashtekar-Krishnan:2004, Dreyer2003,
baumgarteShapiroBook} for a review of the properties of apparent and isolated
horizons. We require boundary conditions on the conformal factor, the shift
vector, and the lapse function.

The boundary condition for the conformal factor is obtained by setting the
expansion scalar on the excision surface to zero, ensuring that it is an
apparent horizon. To see how this results in a boundary condition, we first
write out the expansion of $\mathcal{B}_E$ as
\begin{gather}
\Theta = \frac{4}{\psi^3} \left[ \bar{s}^k~\partial_k \psi
+\frac{\psi^{3}}{8 N} \bar{s}^i \bar{s}^j
\left( (\bar{\Lb} N)_{ij} - \bar{u}_{ij} \right) \right. \nonumber \\
\left. +\frac{\psi}{4} \bar{h}^{ij} \bar{\nabla}_i \bar{s}_j
-\frac{1}{6} K \psi^3 \right],
\label{Eq:Expansion}
\end{gather}
where $\bar{s}^i = \psi^2 s^i$, $s^i$ is the spatial unit normal to
$\mathcal{B}_E$, and $\bar{h}_{ij} = \bar{g}_{ij} - \bar{s}_i \bar{s}_j$
is the induced conformal 2-metric on $\mathcal{B}_E$. $\bar{h}_{ij}$ is
related to the induced 2-metric on $\mathcal{B}_E$ by
$h_{ij} = \psi^4 \bar{h}_{ij}$.
Enforcing the excision surfaces to be apparent horizons (setting $\Theta=0$)
gives us a boundary condition on the conformal factor at $\mathcal{B}_{E}$:
\begin{align}
\bar{s}^k~\partial_k \psi =& - \frac{\psi^{3}}{8 N} \bar{s}^i \bar{s}^j
\left( (\bar{\Lb} N)_{ij} - \bar{u}_{ij} \right) \nonumber \\
 &-\frac{\psi}{4} \bar{h}^{ij} \bar{\nabla}_i \bar{s}_j
+\frac{1}{6} K \psi^3.
\label{Eq:psiHorizonBC}
\end{align}

The boundary condition on the shift is obtained by requiring that:
(1) The coordinate location of the apparent horizons do not change
(in a co-rotating frame)
as the initial data begin to evolve.
(2) The shear tensor vanishes on the excision surface; this is a property
of isolated horizons~\cite{Ashtekar-Krishnan:2004}.
We impose these two conditions only approximately, as described below.
To obtain the shift boundary condition, we first decompose the shift into
parts normal and tangential to the surface $\mathcal{B}_{E}$,
\begin{align}
  N^i =&  ~N^i_{\parallel} + N_{\bot} s^i,
\end{align}
where
\begin{eqnarray}
    N^i_{\parallel} &\equiv& h^i_j N^j, \\
    N_{\bot}       &\equiv& N^i s_i.
\end{eqnarray}

The inner boundary condition (at $\mathcal{B}_{E}$) for the shift is
\begin{eqnarray}
\label{Eq:PerpendicularShiftHorizonBC}
N_{\bot} &=& N, \\
N^i_{\parallel} &=& - \Omega_r^{(k)} \xi^i_{(k)},
\label{Eq:ParallelShiftHorizonBC}
\end{eqnarray}
where
\begin{gather}
  \vec{\xi}_{(0)} = y\hat{z}-z\hat{y},\\
  \vec{\xi}_{(1)} = z\hat{x}-x\hat{z},\\
  \vec{\xi}_{(2)} = x\hat{y}-y\hat{x}
\label{Eq:ConformalKVs}
\end{gather}
are three linearly independent conformal Killing vectors of
a coordinate sphere, and $\Omega_r^{(k)}$ are three arbitrarily
specifiable free parameters that will be discussed below.
The first condition, Eq.~(\ref{Eq:PerpendicularShiftHorizonBC}),
ensures the apparent horizons
are initially at rest in the coordinates.
The second condition, Eq.~(\ref{Eq:ParallelShiftHorizonBC}),
sets the spin of the black hole \cite{Cook2004, Cook2002}.
If the excision surface is a coordinate sphere, then
$\vec{\xi}_{(k)}$ are conformal Killing vectors associated
with $\bar{h}_{ij}$, $\vec{\xi}_{(k)}$ are orthogonal to $s_i$,
and the shear tensor vanishes on the excision surface~\cite{Cook2004}.
For the initial data choices compared here, the excision boundary
is not a coordinate sphere, so neither the shear-free condition
nor the stationary-horizon condition that motivated the shift boundary
conditions are satisfied. Nevertheless, we find that the boundary
conditions above are adequate for binary black hole initial data.

In practice, it is not possible to {\it a priori}
choose values of $\Omega_r^{(k)}$ that will yield a desired
black hole spin; instead one must use an iterative
procedure~\cite{Ossokine:2015yla, Buchman:2012dw}, where at each
iteration $\Omega_r^{(k)}$
is updated until the spin converges to the desired value.  For each
iteration, the spin parameter in the single-black-hole solutions
$\bar{g}^\alpha_{ij}$ and $K^\alpha$
(cf. Eqs.~(\ref{Eq:Superposition}) and~(\ref{Eq:SuperpositionTrK})) is
unchanged, and is set to the desired black hole spin.

Finally, the boundary condition at $\mathcal{B}_{E}$ for the lapse
(which can be chosen freely~\cite{Cook2004}) is chosen such that its value in
the vicinity of each black hole approaches that of the corresponding single
black hole lapse,
\begin{equation}
N \psi = 1 + \sum_{\alpha=1}^2 e^{-r_{\alpha}^2/w_{\alpha}^2} (N_{\alpha} - 1),
\label{Eq:LapseBC}
\end{equation}
where $N_{\alpha}$ is the lapse corresponding to single black hole $\alpha$
and the Gaussian weights are the same as in Eq.~(\ref{Eq:Superposition}).

\begin{figure*}[thb]
\begin{center}
\includegraphics[scale=0.35]{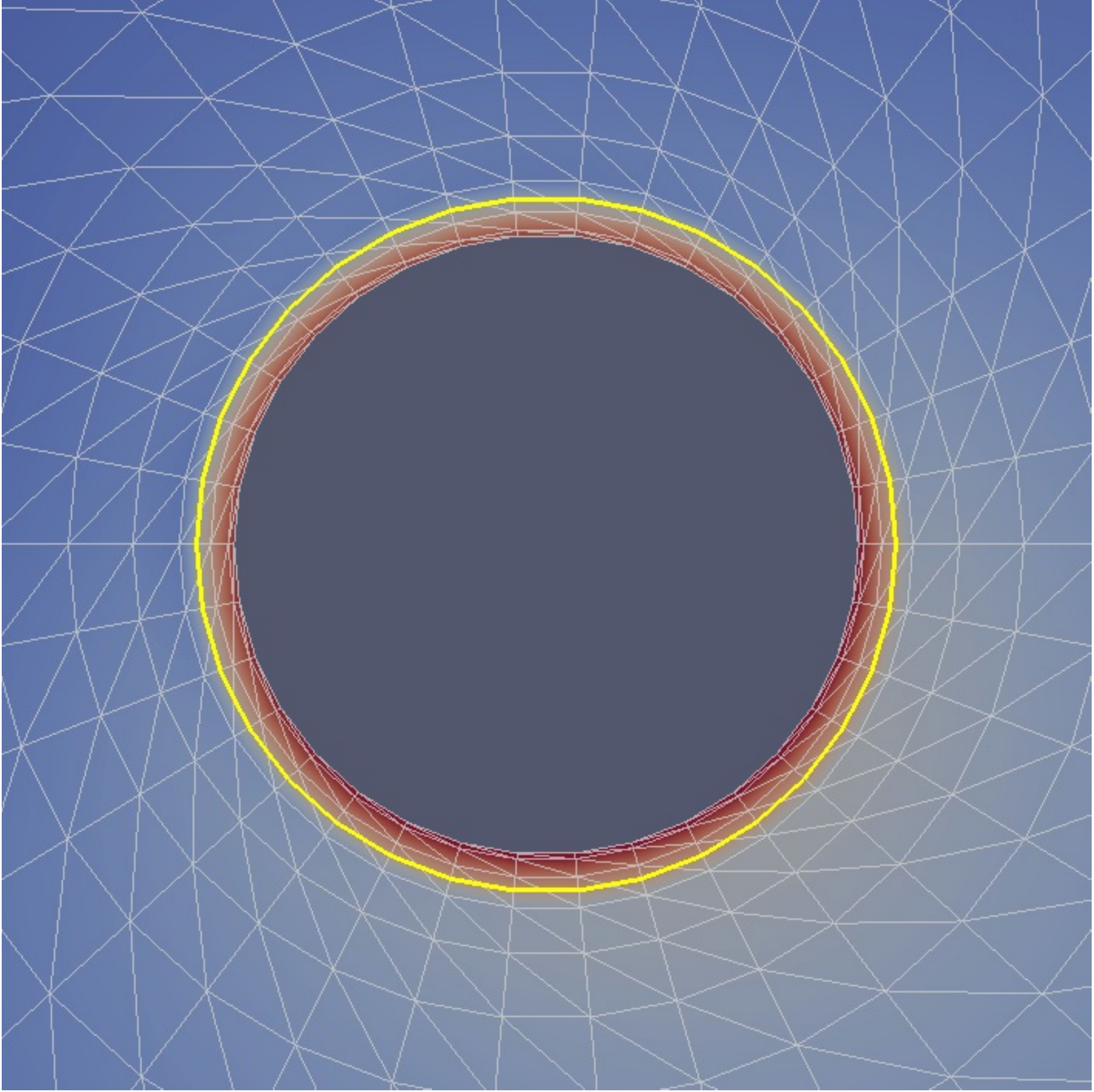} ~~~~
\includegraphics[scale=0.35]{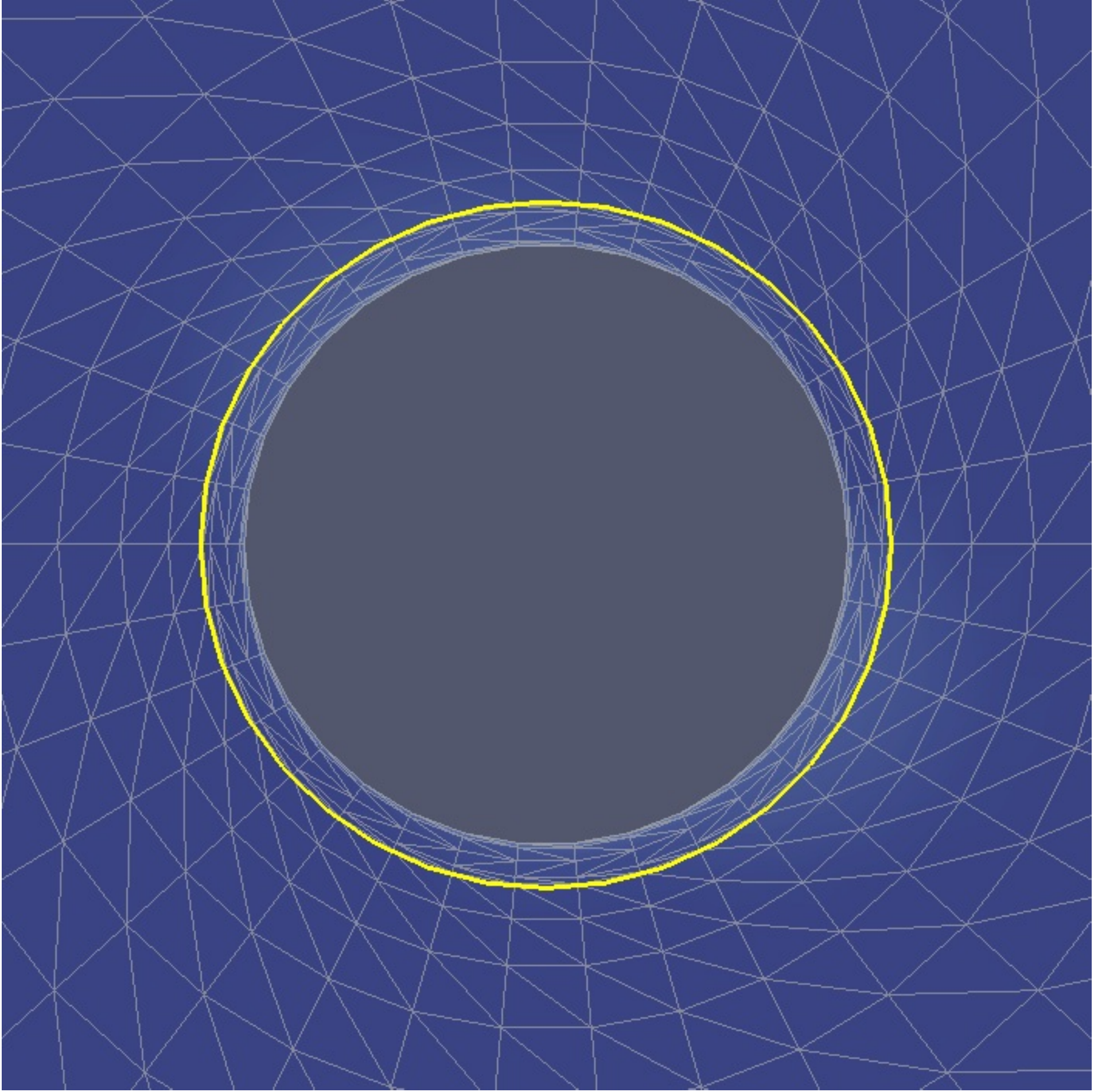} ~~
\includegraphics[scale=0.35]{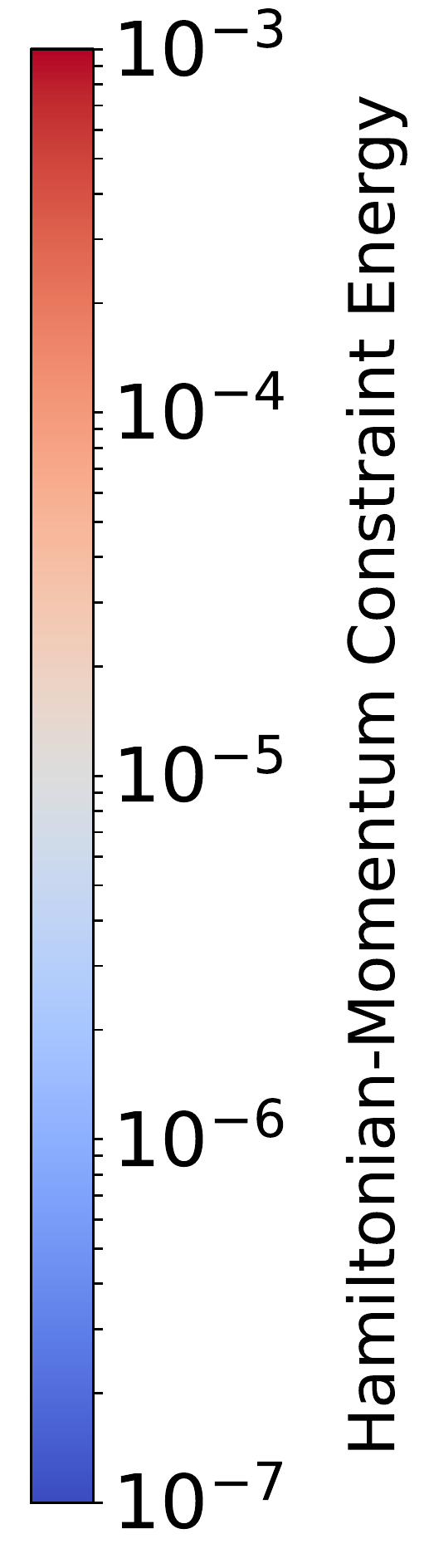}
\caption{
Initial constraint violations on the $z=0$ plane
near the larger hole of a BBH system,
for horizon boundary conditions (left) and negative expansion
boundary conditions (right). Colors show the magnitude of the
Hamiltonian-momentum constraint energy
(cf. Eq.~\ref{Eq:HamMomConstraintEnergy}), the yellow circle is
the apparent horizon, and the large black area inside the
horizon is the excision region.
Here superposed Kerr-Schild free data are used to construct a
BBH with mass ratio $q=1.1$ and spins $\chi_{1z}=-0.3$ and
$\chi_{2z}=-0.4$ along the direction of orbital angular
momentum.  Unlike the horizon boundary conditions, the negative expansion
boundary conditions require no extrapolation inside the horizon, and thus
yield constraints near and inside the apparent horizon that
are about 3 orders of magnitude smaller.
}
\label{Fig:HorPenComparison}
\end{center}
\end{figure*}

\subsubsection{Negative expansion boundary conditions}
\label{Subsubsec:NegExpBC}

The horizon boundary conditions discussed above enforce the
excision surfaces to be apparent horizons.  However, BBH evolutions require an
inner boundary that is slightly {\em inside} the apparent horizons,
for the following reasons:
(1) The apparent horizons dynamically
change shape and size during evolution, so
if the excision surfaces are \emph{at} the
apparent horizons, the horizons
can fall off the numerical grid during evolution.
(2) Our method of finding apparent horizons during the evolution needs to
explore regions just inside and just outside of the horizon in order
to converge onto the correct surface.
(3) During the evolution, no boundary conditions need to be imposed at the
inner boundary, because all characteristic fields are ingoing (into the black
hole) there.  To maintain this ingoing-characteristic-fields condition,
the inner boundary is adjusted to closely track the apparent horizon to
within a small but nonzero error tolerance.

This means that after
solving for initial data using horizon boundary conditions,
the initial data must be extrapolated spatially to a new grid that has
smaller excision surfaces. This extrapolation
introduces constraint violations (cf. left panel of
Fig.~\ref{Fig:HorPenComparison}), and therefore we
propose new boundary conditions that are similar to
the horizon boundary conditions discussed above but are set on a surface
inside the horizon and thus avoid extrapolation altogether.

The idea behind the new boundary conditions is to set the
expansion not to zero, but to some nonzero value that ensures that the
excision boundary is inside an apparent horizon rather than on one.
We use Eq.~(\ref{Eq:Expansion}) to modify the conformal factor boundary
condition at $\mathcal{B}_{E}$ to:
\begin{eqnarray}
\bar{s}^k~\partial_k \psi &=& - \frac{\psi^{3}}{8 N} \bar{s}^i \bar{s}^j
\left( (\bar{\Lb} N)_{ij} - \bar{u}_{ij} \right) \nonumber \\
&&-\frac{\psi}{4} \bar{h}^{ij} \bar{\nabla}_i \bar{s}_j
+\frac{1}{6} K \psi^3 + \frac{\psi^3}{4} \Theta_{\alpha},
\label{Eq:psiHorPenBC}
\end{eqnarray}
where $\alpha$ denotes the particular BH and $\Theta_{\alpha}$ is computed
from the single BH metrics used in Eq.~\ref{Eq:Superposition}.
As we choose the excision surface to be slightly inside the single BH
horizons, $\Theta_{\alpha}$ is negative on the surface. Henceforth we refer
to this boundary condition as a negative expansion boundary condition.

When imposing the negative expansion condition, we also need to modify the
shift boundary condition, as Eq.~(\ref{Eq:PerpendicularShiftHorizonBC})
holds only on a horizon.
Noting that for a single BH, $\epsilon = N_{\bot} - N$
is positive inside the horizon and negative outside, we modify
the boundary condition at $\mathcal{B}_{E}$ for the normal component of shift
to:
\begin{gather}
N_{\bot} = N + \epsilon_{\alpha},
\end{gather}
where $\epsilon_{\alpha} = N_{\bot\alpha} - N_{\alpha}$ are again obtained
from the single BH solutions of the individual holes.

For negative expansion boundary conditions, we continue to use
Eq.~(\ref{Eq:ParallelShiftHorizonBC}) for the tangential part of the
shift. We also continue to use Eq.~(\ref{Eq:LapseBC}) for the boundary
condition on the lapse, with $N_{\alpha}$ evaluated at the new
location of the inner boundary.  We find that the procedure for
setting the spin via iteration over $\Omega_r^{(k)}$, as described in
Sec.~\ref{Subsubsec:HorBC}, works just as well in the case of a negative
expansion BC as it does for a horizon BC.

Figure~\ref{Fig:HorPenComparison} demonstrates the efficacy of these new
boundary conditions; shown are the constraints near the larger black hole
when using horizon boundary conditions and the new negative expansion boundary
conditions. When using negative expansion boundary conditions, the constraints
improve by about 3 orders of magnitude inside and near the apparent horizon.
Note, however, that once the evolution begins, most of this constraint
violation propagates inwards into the excision surfaces and out of the
computational domain. This is because in the generalized harmonic formalism
the evolution of constraint violations is governed by a wave
equation~\cite{Lindblom:2007}, which ensures that constraint violations
propagate causally. Hence, we do not expect the new boundary conditions to
reduce constraint violations during the evolution nearly as much as they
improve initial constraint violations.

\subsection{Gauge choices}
\label{Subsec:IDGaugeChoices}

SpEC uses the generalized harmonic evolution system~\cite{Lindblom:2007,
Friedrich1985, Garfinkle2002, Pretorius2005c} to evolve the initial data. In
this formalism, the gauge choice is set by requiring the coordinates to satisfy
an inhomogeneous wave equation,
\begin{equation}
-{^{(4)}}\Gamma^a = \nabla^c \nabla_c x^a = \h^a,
\label{Eq:GaugeCondition}
\end{equation}
where $^{(4)}\Gamma^a=\psi^{bc} \, {^{(4)}}\Gamma^a_{bc}$, $\psi_{ab}$ is the
spacetime
metric, $^{(4)}\Gamma^a_{bc}$ are the Christoffel symbols associated with
$\psi_{ab}$, $\nabla_a$ is the covariant derivative operator compatible with
$\psi_{ab}$, and $\h^a$ (called the gauge source function) is a function of
the coordinates $x^a$ and the metric $\psi_{ab}$ (but not the derivatives of
the metric).

The simplest choice for the gauge source function is to set it to zero, which
yields the harmonic gauge:
\begin{equation}
\label{Eq:HarmGaugeCondition}
\nabla^c \nabla_c x^a = \h^a =  0.
\end{equation}
Harmonic coordinates have proven to
be extremely useful in analytic studies in GR
\cite{deDonder1921,Lanczos22,Choquet1952,fischer_marsden72,cook_scheel97}.
However, this gauge does not work
well for simulations of black hole mergers. One common reason for the failure
is growth in $\sqrt{g}/N$, which tends to blow up as the black holes approach
each other~\cite{ Szilagyi:2009qz}.

SpEC evolutions are done instead in the damped harmonic
gauge~\cite{Szilagyi:2009qz} given by:
\begin{gather}
\label{Eq:DHGaugeCondition}
\nabla^c \nabla_c x^a = \h^a_{DH}, \\
\h^a_{DH} \equiv \mu_L \log\left(\frac{\sqrt{g}}{N}\right) t^a
-\mu_S \frac{N^i}{N} g^a{}_{i},
\label{Eq:HDHLowerindex}
\end{gather}
where $t^a$ is the future directed unit normal
to constant-t hypersurfaces, $g_{ab}$ is the spatial metric of the constant-$t$
hypersurfaces and $g$ its determinant, and $\mu_L$ and $\mu_S$
are positive damping factors that can be chosen arbitrarily.
The spatial coordinates and lapse satisfy a damped wave equation
with damping factors $\mu_S$ and $\mu_L$, and are driven
towards solutions of the covariant spatial Laplace equation
on timescales of $1/\mu_S$ and $1/\mu_L$, respectively.
Damped harmonic gauge tends to reduce extraneous gauge dynamics present in the
harmonic gauge.

The damping factors are chosen as follows:
\begin{equation}
\label{Eq:Mu_definition}
\mu_S = \mu_L = \mu_0~\left[\log\left(\frac{\sqrt{g}}{N}\right) \right]^2 ,
\end{equation}
where $M \mu_0$ is chosen to be of order unity, and $\mu_0$ is a function of
time (to accommodate starting an evolution from initial data satisfying a
different gauge condition).  This choice of the damping factors ensures that
$\sqrt{g}/N$ is driven faster than exponentially towards an asymptotic
state~\cite{Szilagyi:2009qz}, so that $\sqrt{g}/N$ does not grow rapidly near
mergers as often happens with harmonic gauge.

\begin{center}
\begin{table*}
\begin{tabular}{||c |l |c |l |l ||}
\hline
\multirow{2}{*}{Type} & \multicolumn{2}{c|}{Initial Data} & \multicolumn{2}{c||}{Evolution} \\
\cline{2-5}
& \multicolumn{1}{c|}{XCTS free data ($\bar{g}_{ij}$, $K$) }
& \multicolumn{1}{c|}{Inner BC }
& \multicolumn{1}{c|}{Initial Gauge}
& \multicolumn{1}{c||}{Final Gauge }
\\ [0.5ex]
\hline
\cline{1-5}
\SKSHorBC  & Superposed Kerr-Schild     & Horizon BC   & Quasi-equilibrium  & Damped Harmonic \\
\hline
\SKS       & Superposed Kerr-Schild     & Negative expansion BC & Quasi-equilibrium  & Damped Harmonic \\
\hline
\SHK       & Superposed Harmonic   & Negative expansion BC & Harmonic   & Damped Harmonic \\
\hline
\SDH       & Superposed Damped Harmonic & Negative expansion BC & Damped Harmonic      & Damped Harmonic \\
\hline
\SKSDH & Superposed Kerr-Schild     & Negative expansion BC & Damped Harmonic      & Damped Harmonic \\
\hline
\end{tabular}
\caption{Types of initial data considered in this study. The initial data
formalism is described in Sec.~\ref{Sec:IDFormalism}. See
Sec.~\ref{Subsec:XCTS} for the XCTS system of equations and
Sec.~\ref{Subsubsec:XCTSFreeData} for the freely specifiable data in XCTS.
We describe the horizon boundary conditions in Sec.~\ref{Subsubsec:HorBC} and
negative expansion boundary conditions in Sec.~\ref{Subsubsec:NegExpBC}. The
gauge choices are described in Sec.~\ref{Subsec:IDGaugeChoices}. The initial
gauge is chosen by setting $\partial_t N$ and $\partial_t N^i$ according to
Sec.~\ref{Subsubsec:dtLapseShift}.
}
\label{Tab:IDTypes}
\end{table*}
\end{center}

\subsubsection{Setting the initial gauge}
\label{Subsubsec:dtLapseShift}

The generalized harmonic evolution system requires the metric $\psi_{ab}$ and
its time derivative $\partial_t \psi_{ab}$ to be specified on the initial time
slice. Most of these quantities are determined by the solution of the XCTS
equations and the free data that are used in solving these equations.  However,
$\partial_t \psi_{ab}$ also includes the time derivatives of the lapse
and shift, which are independent of
the XCTS equations. Instead, they are equivalent to the
initial choice of the gauge source function $\h^a$.  To see this, we
expand the generalized harmonic gauge condition,
Eq.~(\ref{Eq:GaugeCondition}), and rewrite it in terms of the time
derivatives of lapse and shift:
\begin{eqnarray}
\label{Eq:dtLapse}
\partial_t N &=& N^j \partial_{j}N -N^2 K  +N^3 \h^0 , \\
\partial_t N^i &=& N^j \partial_{j}N^i -N^2 g^{ij} \partial_{j}(\log N)
+N^2 \Gamma^i \nonumber \\
&&+ N^2 (\h^i+ N^i \h^0) .
\label{Eq:dtShift}
\end{eqnarray}
Here $\Gamma^i=g^{jk} \Gamma^i_{jk}$ and $\Gamma^i_{jk}$ are the Christoffel
symbols associated with $g_{ij}$. Note that $N^2$ and $N^3$ indicates powers
of the lapse function, whereas $N^i$, $\h^0$ and $\h^i$ are components of the
shift-vector $N^i$ and the gauge-source function $H^a$.

The default choice in SpEC simulations has been to set $\partial_t N=\partial_t
N^i=0$ in a frame co-rotating with the binary; this is meant to be a
quasiequilibrium condition that reduces initial gauge dynamics.
Given this choice, Eqs.~(\ref{Eq:dtLapse}) and~(\ref{Eq:dtShift}) determine the
initial values of $H^a$, which are kept time-independent in
this co-rotating frame during the initial stages of the evolution.
However, the damped harmonic gauge works best for mergers, so SpEC
simulations customarily move from co-rotating gauge to
damped harmonic gauge via a
smooth gauge transformation during the first $\sim 50 M$ of the
evolution.
However, this gauge transformation introduces additional
complications:
(1) The gauge change causes additional gauge
dynamics in the evolution.
(2) The gauge change happens at the same time as the junk radiation leaves the
system, making it difficult to distinguish junk radiation from gauge dynamics.
(3) The gauge change impacts the ability to achieve configurations with zero
orbital eccentricity. To understand this last point, we note that SpEC
evolutions customarily employ iterative eccentricity
reduction~\cite{Buonanno:2010yk}: Starting with orbital parameters
predicted by post-Newtonian theory, we evolve the binary for $\sim2$ orbits,
compute the eccentricity, adjust the initial parameters and repeat until the
desired eccentricity is achieved. This involves an extrapolation back in
time to compute adjusted parameters and this extrapolation
happens at the same time as the gauge transformation.

\subsubsection{New choices of initial gauge}

With the aim of addressing these issues, as part of this work we have
also explored setting the initial gauge to satisfy the harmonic or
damped harmonic condition, as explained in more detail in
Sec.~\ref{Sec:IDTypes}.  In order to set the initial gauge to the
harmonic or damped harmonic gauge, we set $\partial_t N$ and
$\partial_t N^i$ according to Eqs.~(\ref{Eq:dtLapse})
and~(\ref{Eq:dtShift}) at $t=0$, with $\h^a=0$ for harmonic gauge and
$\h^a=\h^a_{DH}$ for damped harmonic gauge.

\section{BBH Initial Data Types}
\label{Sec:IDTypes}

Having introduced the BBH initial data formalism, in this section we discuss
the different initial data sets considered in this study; these are also
listed in Table~\ref{Tab:IDTypes}.
Our naming convention for the
initial data sets indicates the choice of free data, initial gauge condition
and boundary conditions at excision surfaces.
For example, \SKSHorBC stands for superposed Kerr-Schild free data,
quasi-equilibrium initial gauge condition, and horizon boundary conditions at
excision surfaces. Unless explicitly specified, we use the new negative
expansion boundary conditions at excision surfaces.

\subsection{Superposed Kerr-Schild with horizon boundary conditions (\SKSHorBC)}
This is the type of initial data currently implemented in
SpEC~\cite{Lovelace2008}. Initial data are constructed by solving the XCTS
system of equations, with horizon boundary conditions imposed on the excision
surfaces. The free data for XCTS equations are obtained using a superposition
of two single BHs in the Kerr-Schild gauge. Once the XCTS equations are solved,
the initial data are extrapolated slightly inside the apparent horizon surfaces.
The initial gauge is set by imposing $\partial_t N = \partial_t N^i=0$ in a
co-rotating frame. During the initial stages of the evolution
a smooth gauge transformation moves
into the damped harmonic gauge over a time scale
of $50M$.  We refer to this initial data set as \SKSHorBC.

\subsection{Superposed Kerr-Schild with negative expansion boundary
conditions (\SKS)}

This is the same as \SKSHorBC above but with a negative expansion boundary
condition (Sec.~\ref{Subsubsec:NegExpBC}) on the excision surfaces.
We choose the excision surfaces to be slightly inside the
apparent horizons and thus avoid the need for extrapolation in initial data.
We refer to this as \SKS.

\subsection{Superposed Harmonic-Kerr (\SHK)}
The free data are obtained by superposing two single BHs in
the harmonic coordinates of Ref.~\cite{cook_scheel97}.
The time derivatives
$\partial_t N$ and $\partial_t N^i$ at $t=0$ are set according
to the Harmonic gauge condition (cf. Eqs.~\ref{Eq:HarmGaugeCondition},
\ref{Eq:dtLapse} and \ref{Eq:dtShift}):
\begin{gather}
\partial_t N = N^j \partial_{j}N -N^2 K,  \\
\partial_t N^i = N^j \partial_{j}N^i -N^2 g^{ij} \partial_{j}(\log N)
+N^2 \Gamma^i.
\end{gather}

Therefore, the initial data is in the harmonic gauge at $t=0$.
As in the case of \SKS, during the initial stages of the evolution we do a
smooth gauge transformation to the damped harmonic gauge over a time scale of
$50M$. A negative expansion boundary condition
(Sec.~\ref{Subsubsec:NegExpBC}) is used on the excision surfaces.
We refer to this initial data as \SHK. We find that \SHK initial data
works well for dimensionless spin magnitudes $\chi \leq 0.7$; for higher spins
the single BHs in harmonic coordinates are highly compressed
in the direction of spin (see Fig.~\ref{Fig:Pancake}).

\subsection{Superposed Damped Harmonic (\SDH)}
The free data are obtained by superposing two single BHs in
the damped harmonic gauge of Ref.~\cite{Varma:2018sbhdh}, and a
negative expansion boundary condition (Sec.~\ref{Subsubsec:NegExpBC}) is used
on the excision surfaces.
$\partial_t N$ and $\partial_t N^i$ at $t=0$ are set according
to the damped harmonic gauge condition, Eqs.~(\ref{Eq:HDHLowerindex}),
(\ref{Eq:dtLapse}) and (\ref{Eq:dtShift}):
\begin{eqnarray}
\label{Eq:DHSetdtLapse}
\partial_t N &=& N^j \partial_{j}N -N^2 K  +N^3 \h^0_{DH},  \\
\partial_t N^i &=& ~N^j \partial_{j}N^i -N^2 g^{ij} \partial_{j}(\log N)
+N^2 \Gamma^i \nonumber \\
&&+ N^2 (\h^i_{DH}+ N^i \h^0_{DH}).
\label{Eq:DHSetdtShift}
\end{eqnarray}
Because the initial data are already in the
damped harmonic gauge at $t=0$, no
gauge transformation is necessary during the evolution.
We refer to this initial data set as \SDH.

\subsection{Superposed Kerr-Schild with Damped Harmonic Gauge (\SKSDH)}

This is the same as \SKS, except the initial gauge
is set to the damped harmonic gauge using
Eqs.~(\ref{Eq:DHSetdtLapse}) and (\ref{Eq:DHSetdtShift}).
Because the damped harmonic gauge condition is
satisfied at $t=0$, no gauge
transformation is needed during evolution. We refer to these initial
data as \SKSDH. Although the motivation for \SKSDH is to avoid the smooth
gauge transformation during the evolution, for \SKSDH the gauge
is not in quasi-equilibrium at $t=0$ even if the BHs are
far apart; this could potentially
lead to more gauge dynamics at the start of the evolution.

\section{Convergence of initial data}
\label{Sec:IDTests}

In this section, we perform a convergence test of the different initial data
sets we construct. We use the spectral elliptic solver described in
Refs.~\cite{Pfeiffer2003, Ossokine:2015yla} to solve the XCTS equations. We
compare the Hamiltonian and momentum constraint violations at different
resolutions, for the case of a nonprecessing BBH system with mass ratio $q=1.1$
and dimensionless spins $\chi_{1z}=-0.3$, $\chi_{2z}=-0.4$ along the orbital
angular momentum direction.  The Hamiltonian and momentum constraints in vacuum
are given by:
\begin{gather}
R + K^2 - K_{ij} K^{ij} = 0, \\
g^{jk} (\nabla_j K_{ki} - \nabla_i K_{jk}) = 0,
\end{gather}
where $R$ and $\nabla_i$ are the Ricci scalar and  the spatial covariant
derivative operator associated with $g_{ij}$. We quantify these constraint
violations by computing their $L^2$ norms over the initial data domain. We
also normalize them to obtain dimensionless quantities
\footnote{Notice that for the
denominator of
Eqs.~(\ref{Eq:HamConstraintEnergy}) and (\ref{Eq:MomConstraintEnergy}) as
well as Eq.~(\ref{Eq:DHConstraintEnergyVector}) below,
repeated indices are summed over \emph{after} squaring the quantities, unlike
the standard summation notation.},
\begin{gather}
  \label{Eq:HamConstraintEnergy}
\mathcal{H} =\!\!  \frac{\Vert R + K^2 - K_{ij} K^{ij} \Vert}
{ \left \Vert \sqrt{  \sum\limits_{i,j,k} \!\left[
(R_{ij}g^{ij})^2
\!\!+\!\!(K_{ij}K_{kl} g^{ik}g^{jl})^2
\!\!+\!\!(K_{ij}K_{kl} g^{ij}g^{kl})^2  \right]
} ~ \right \Vert }, \\
  \label{Eq:MomConstraintEnergy}
\mathcal{M}_i = \frac{\Vert g^{jk} (\nabla_j K_{ki} - \nabla_i K_{jk})\Vert}
{\left \Vert\sqrt{\sum\limits_{i,j,k} \Big[ (g^{jk})^2 ((\nabla_j K_{ki})
^2 + (\nabla_i K_{jk})^2) \Big] } ~\right\Vert},
\end{gather}
where $\Vert . \Vert$ denotes the $L^2$ norm over the domain. Finally, we
define a Hamiltonian-Momentum constraint energy:
\begin{gather}
\mathcal{C} = \sqrt{\mathcal{H}^2 + \sum_{i=0}^2 \mathcal{M}_i^2} \,.
\label{Eq:HamMomConstraintEnergy}
\end{gather}

\begin{figure}[hbt]
\begin{center}
\includegraphics[scale=0.5]{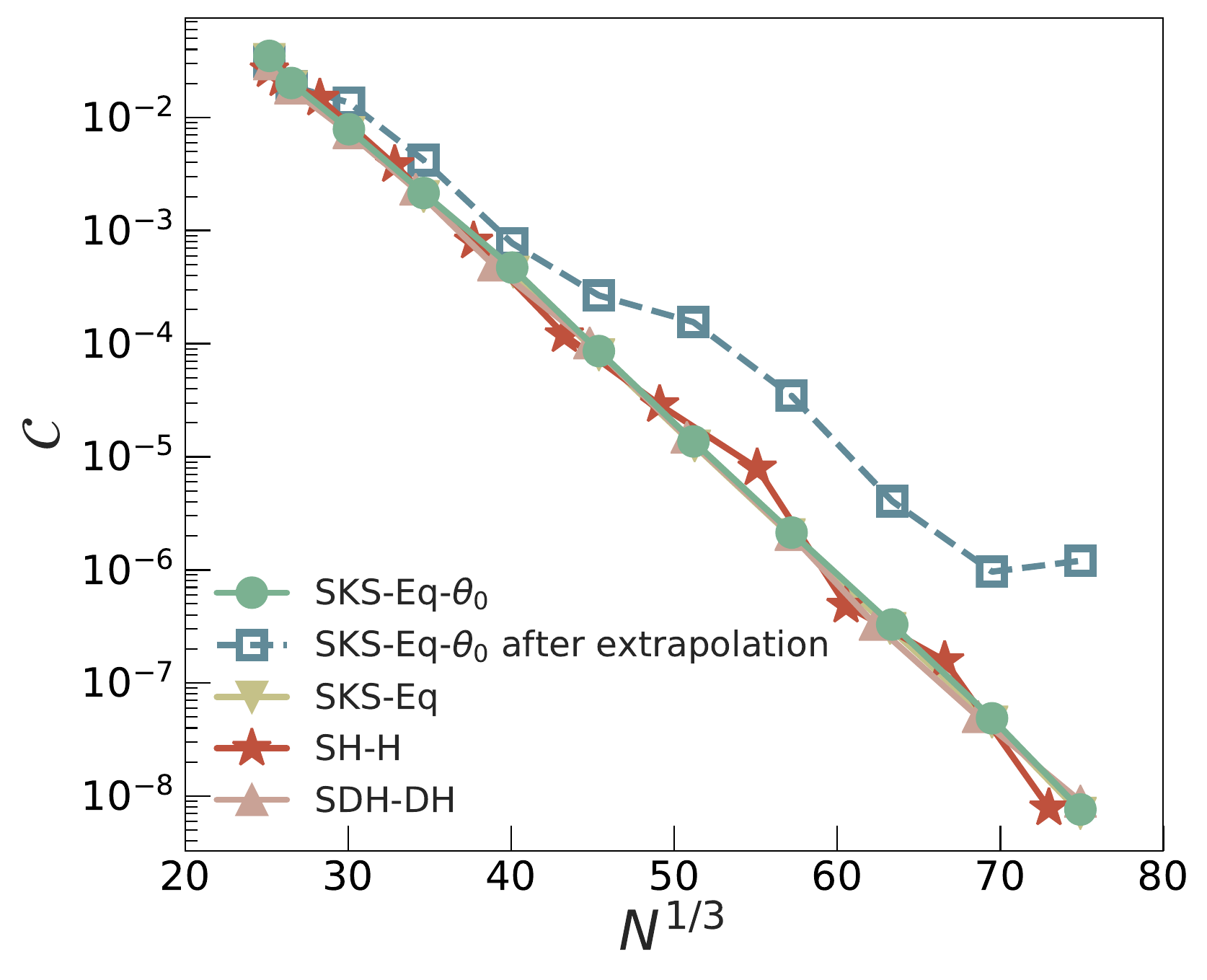}
\caption{Convergence test for the spectral elliptic solver in solving the
XCTS equations for the different initial data types listed in
Table.~\ref{Tab:IDTypes}. Shown is
the Hamiltonian-momentum constraint energy
(Eq.~\ref{Eq:HamMomConstraintEnergy}) vs. the number
of collocation points per dimension in the domain.
As expected for spectral methods, the constraints decrease exponentially.
Also shown are the constraints for \SKSHorBC after extrapolation of
initial data,
where, at high resolution, the constraint violation from extrapolation
dominates (cf. Fig.~\ref{Fig:HorPenComparison}).
There is no extrapolation for \SKS, \SHK, and \SDH, as we
use negative expansion boundary conditions for these.
Note that \SKSDH is not shown here because its solution
of the XCTS equations is identical to \SKS; the cases \SKS
and \SKSDH differ only in the initial gauge condition.
}
\label{Fig:bbh_conv}
\end{center}
\end{figure}

Figure~\ref{Fig:bbh_conv} shows a convergence test for the different initial
data sets considered in this study. We see exponential convergence in all
cases, as is expected with spectral methods. For \SKSHorBC, while we see
exponential convergence for the constraints before extrapolation, the
constraints after extrapolation are significantly higher. This is why we
introduced the new negative expansion boundary condition, which avoids
extrapolation by placing the  excision surface inside rather than at the
apparent horizons.

\section{BBH evolution with different initial data sets}
\label{Sec:BBhEv}

In this section we evolve the different initial data sets discussed above and
compare them for a nonprecessing BBH system with mass ratio
$q=1.1$ and dimensionless spins $\chi_{1z}=-0.3$, $\chi_{2z}=-0.4$
along the orbital angular momentum direction. In particular we look at the
constraint violations, gauge evolution, component parameters, extracted
waveforms, junk radiation, simulation expense, and ease of constructing
zero-eccentricity initial data.

\begin{figure}[hbt]
\begin{center}
\includegraphics[scale=0.6]{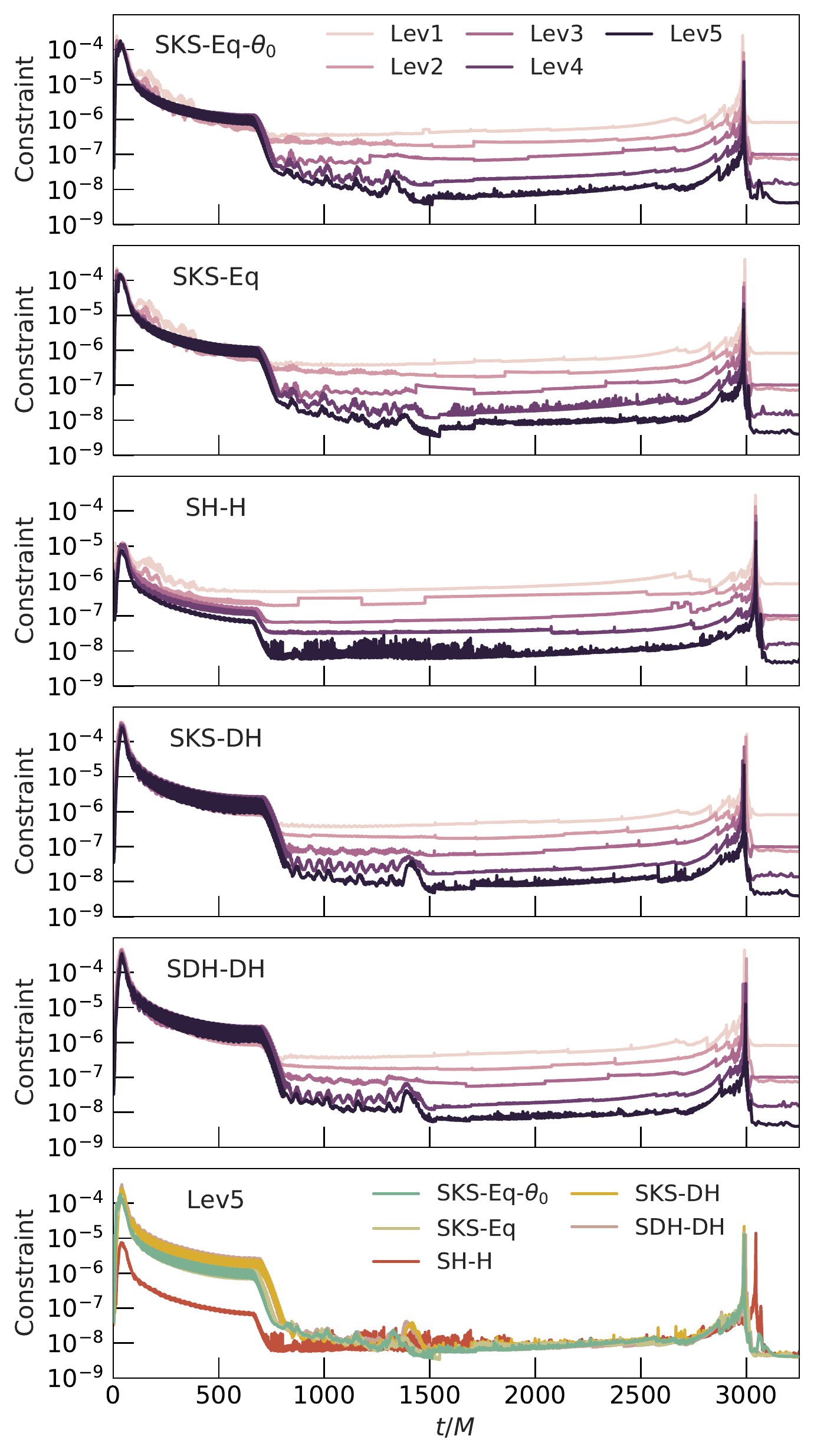}
\caption{Convergence test for constraints during evolution using different
initial data sets. The top panels show the constraints for different
resolutions for each case: Lev1 corresponds to the lowest resolution and Lev5
corresponds to the highest resolution. After the junk radiation leaves the
system, we see convergence in all cases. However, we get good convergence
during junk radiation stages only for \SHK. The bottom panel shows the
constraints for the highest resolution for each case. We see that for \SHK,
the constraints during junk radiation are smaller by about an order of
magnitude.
}
\label{Fig:BBHEvConvTest}
\end{center}
\end{figure}

We performed each of these simulations for 5 different
resolutions in order to do a convergence study. Each resolution is determined
by specifying an error tolerance to our adaptive mesh refinement (AMR)
algorithm~\cite{Szilagyi:2014fna}. In order to match this error tolerance as
the evolution proceeds, AMR adds or removes collocation points from each
subdomain (p-type refinement) and also splits a single subdomain into two or
joins two neighboring subdomains as needed (h-type refinement). We use the
labels ``Lev1'' through ``Lev5'' to indicate decreasing values of AMR error
tolerance. During the junk radiation stage, we intentionally prevent the AMR
algorithm~\cite{Szilagyi:2014fna} from resolving the high-frequency features
present in the initial transients. This
is done because attempting to resolve these
features slows down the evolution considerably, and for most purposes (such as
comparing with LIGO data) the junk-containing part of the waveforms is removed
anyway.

\subsection{Constraint violations}
\label{Subsec:EvConstraints}

Figure~\ref{Fig:BBHEvConvTest} shows the generalized harmonic constraint energy
(defined in Eq.(53) of Ref.~\cite{Lindblom:2007}) during the evolution of the
initial data sets for different resolutions.  As expected, we see convergence
for all the cases after the junk radiation has left the system.  Because we
intentionally prevent the AMR algorithm from resolving the high-frequency
junk-radiation features, it is no surprise that we lose exponential convergence
during the junk stage ($t \lesssim 700M$) for most of the cases considered.
However, for \SHK initial data, we still retain exponential convergence for
most of the junk stage, i.e. for $100M\lesssim t\lesssim 700M$, although with a
shallower slope than at later times.  This indicates that that there are less
prominent high-frequency features present during the junk for \SHK initial
data. The bottom panel of Fig.~\ref{Fig:BBHEvConvTest} shows the constraints
for the highest resolution for different initial data sets. We see that during
the initial junk radiation stage, the constraints are lower for \SHK by about
an order of magnitude compared to \SKSHorBC. \SDH and \SKSDH initial data sets
result in slightly higher constraint violations during junk radiation than
\SKSHorBC, but not by much.

\subsection{Approach to damped harmonic gauge}
\label{Subsec:DHConstriaintTest}

The evolution of each initial data set discussed above eventually
settles into damped harmonic gauge (Eq.~\ref{Eq:GaugeCondition}).  For
\SDH and \SKSDH, the initial data should already be in damped harmonic
gauge, and for the other cases damped harmonic gauge is achieved via
an explicit gauge transformation.  Here we quantify to what extent the
evolutions of these initial data sets actually satisfy
the damped harmonic gauge condition.  Using
Eqs.~(\ref{Eq:GaugeCondition}) and (\ref{Eq:DHGaugeCondition}), we
define a normalized damped harmonic constraint energy,
\begin{gather}
\label{Eq:DHConstraintEnergy}
\mathcal{C}_{DH} =
    \sqrt{\sum_{a=0}^3 \mathcal{C}^a_{DH} \mathcal{C}^a_{DH}} \,,
\end{gather}
\begin{gather}
\label{Eq:DHConstraintEnergyVector}
\mathcal{C}^a_{DH} = \frac{\Vert {^{(4)}}\Gamma^a + \h^a_{DH} \Vert}
{\left\Vert
\sqrt{\sum\limits_{a,b,c=0}^{3} \Big[ (\psi^{bc}\,{^{(4)}}\Gamma^a_{bc})^2
+ (\h^a_{DH})^2 \Big] } ~\right \Vert},
\end{gather}
where $\Vert . \Vert$ denotes the $L^2$ norm over the domain.  We call this
quantity an ``energy'' because it represents one piece of the constraint energy
defined in Eq.~(53) of Ref.~\cite{Lindblom:2007}.

\begin{figure}[bht]
\begin{center}
\includegraphics[scale=0.59]{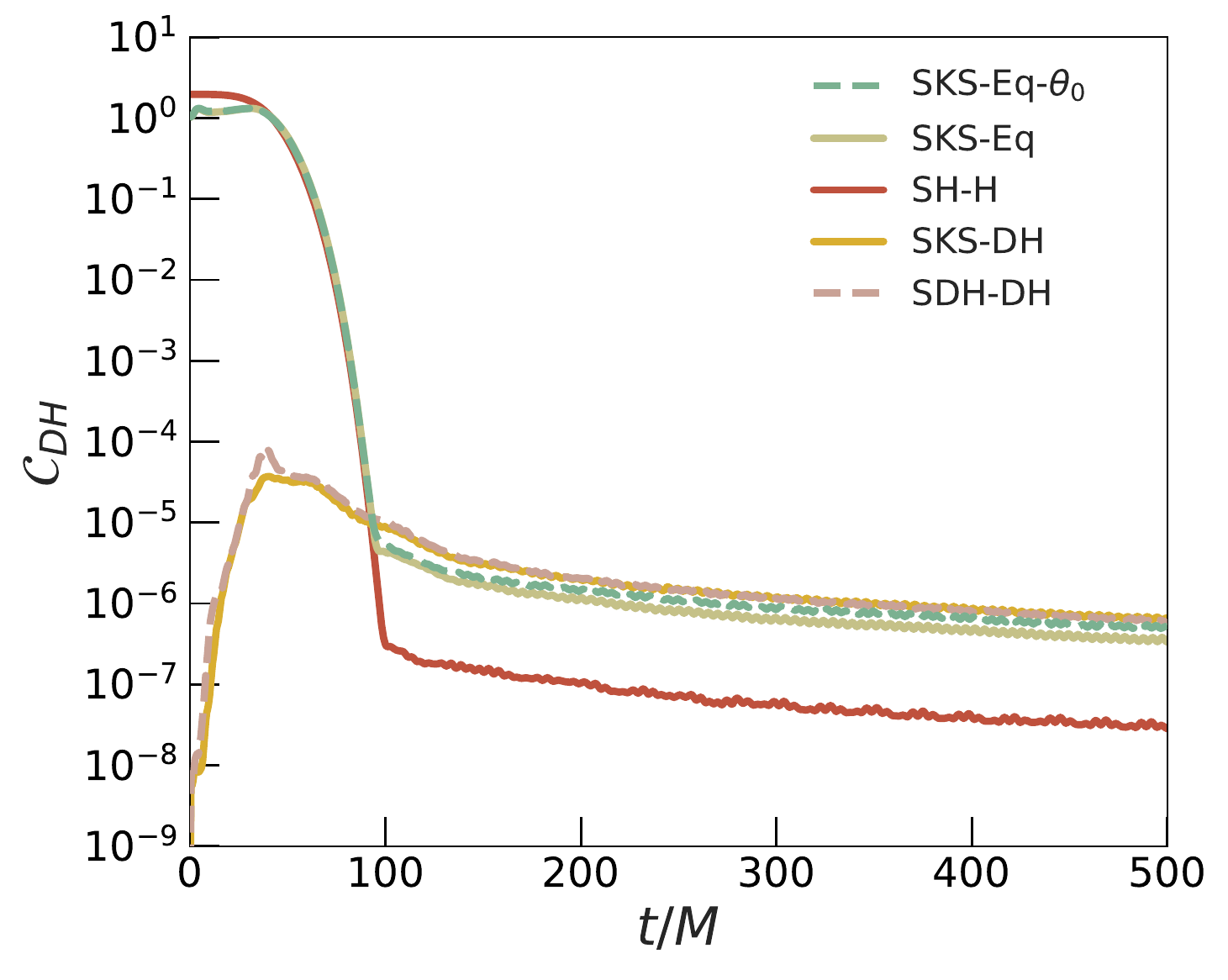}
\caption{Damped harmonic constraint energy (Eq.~\ref{Eq:DHConstraintEnergy})
during evolution of different initial data sets. The damped harmonic constraint
energy quantifies to what extent the gauge satisfies the damped harmonic
condition. For \SDH and \SKSDH initial data sets, the
initial data are already in the damped harmonic gauge.
For the other cases, a smooth gauge transformation is done during early
evolution, on a time scale of about $50 M$, to move into the damped harmonic
gauge. The curves for \SKSHorBC and \SKS lie nearly on top of each other.
\label{Fig:DHEvConstraints}
}
\end{center}
\end{figure}

Figure~\ref{Fig:DHEvConstraints} shows the damped harmonic constraint
energy during evolution of different initial data sets.
For \SDH and \SKSDH, since initial data are already in the damped harmonic
gauge, $\mathcal{C}_{DH}$ starts at about $10^{-8}$, and rises during the junk
radiation stage. However, $\mathcal{C}_{DH}$ always stays
below about $10^{-4}$. Furthermore, the two methods to generate damped
harmonic initial data give rise to comparable $\mathcal{C}_{DH}$.
We find that this peak value of $10^{-4}$ does not change
significantly with resolution. This is understandable, as this is caused by
junk radiation, which we intentionally do not fully resolve.
\SKSHorBC, \SKS, and \SHK start in a
different gauge, and there is no reason to expect small $\mathcal{C}_{DH}$ at
$t=0$. $\mathcal{C}_{DH}$ falls as the evolution transitions to damped
harmonic gauge around $t\sim50M$. The damped harmonic constraint
values after the gauge transformation are
lower for \SHK than for all the other cases because of smaller
junk radiation content,
as we will see in Sec.\ref{Subsec:WaveformComparison} below.

\subsection{Component parameters}

\begin{figure}[bht]
\begin{center}
\includegraphics[scale=0.52]{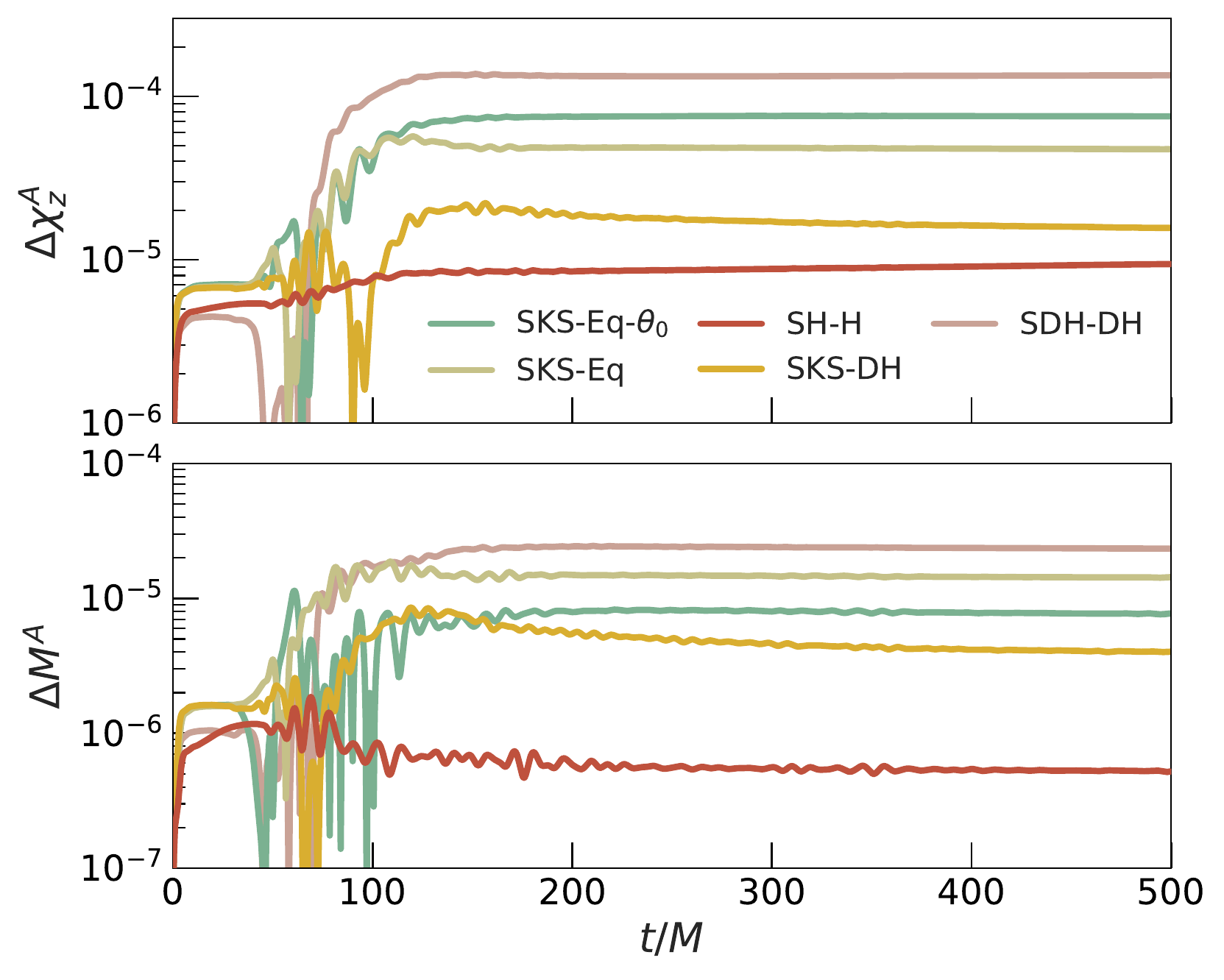}
\caption{Behavior of dimensionless spin along
the angular momentum direction (top panel) and mass (bottom panel) of the
larger black hole during the initial stages of the evolution.
Here, $\Delta M^{A} = |M^{A}(t) - M^{A}(t=0)|$ and
$\Delta \chi^{A}_z = |\chi^{A}_z(t) - \chi^{A}_z(t=0)|$.
The mass and spin are much more stable for \SHK than
for the other cases.  We attribute this to the small amount of junk
radiation in this case; see Sec.~\ref{Subsec:WaveformComparison} below.
}
\label{Fig:CompParams}
\end{center}
\end{figure}

At the start of the evolution, the component spins and masses change slightly
with time. This typically results in slightly lower spins than what we start
with. These changes occur as a result of initial transients such as junk
radiation leaving the system. Note also that in our initial data we do not
tidally deform the BHs.  Hence, the initial component parameters can change as
the BHs settle down into their equilibrium shapes.  Figure~\ref{Fig:CompParams}
shows the change in mass and spin of the larger black hole (with respect to the
simulation input parameters), as the simulation progress. We see that the
component parameters are more stable by about an order of magnitude for the
\SHK initial data compared to \SKSHorBC.  \SDH initial data results in the
largest changes while \SKSDH does better than \SKSHorBC.  In
Sec.~\ref{Subsec:WaveformComparison} we will see that this can be attributed to
the amount of junk radiation for each of these initial data sets.  Note that
Fig.~\ref{Fig:CompParams} corresponds to the highest resolution (Lev=5) used
for this study.  Repeating Fig.~\ref{Fig:CompParams} with a lower resolution
results in changes on the order of $10^{-4}$ in spin and $10^{-5}$ in mass for
all cases except \SHK, and changes on the order of $10^{-5}$ in spin and
$10^{-6}$ in mass for \SHK.  Since the changes with resolution are on the same
order as the variations shown in the figure, the curves in
Fig.~\ref{Fig:CompParams} should be regarded only as order of magnitude
estimates.  For all resolutions, the variations in mass and spin for \SHK are
smaller than for the other cases.

\begin{figure*}[tbh]
\begin{center}
\includegraphics[scale=0.62]{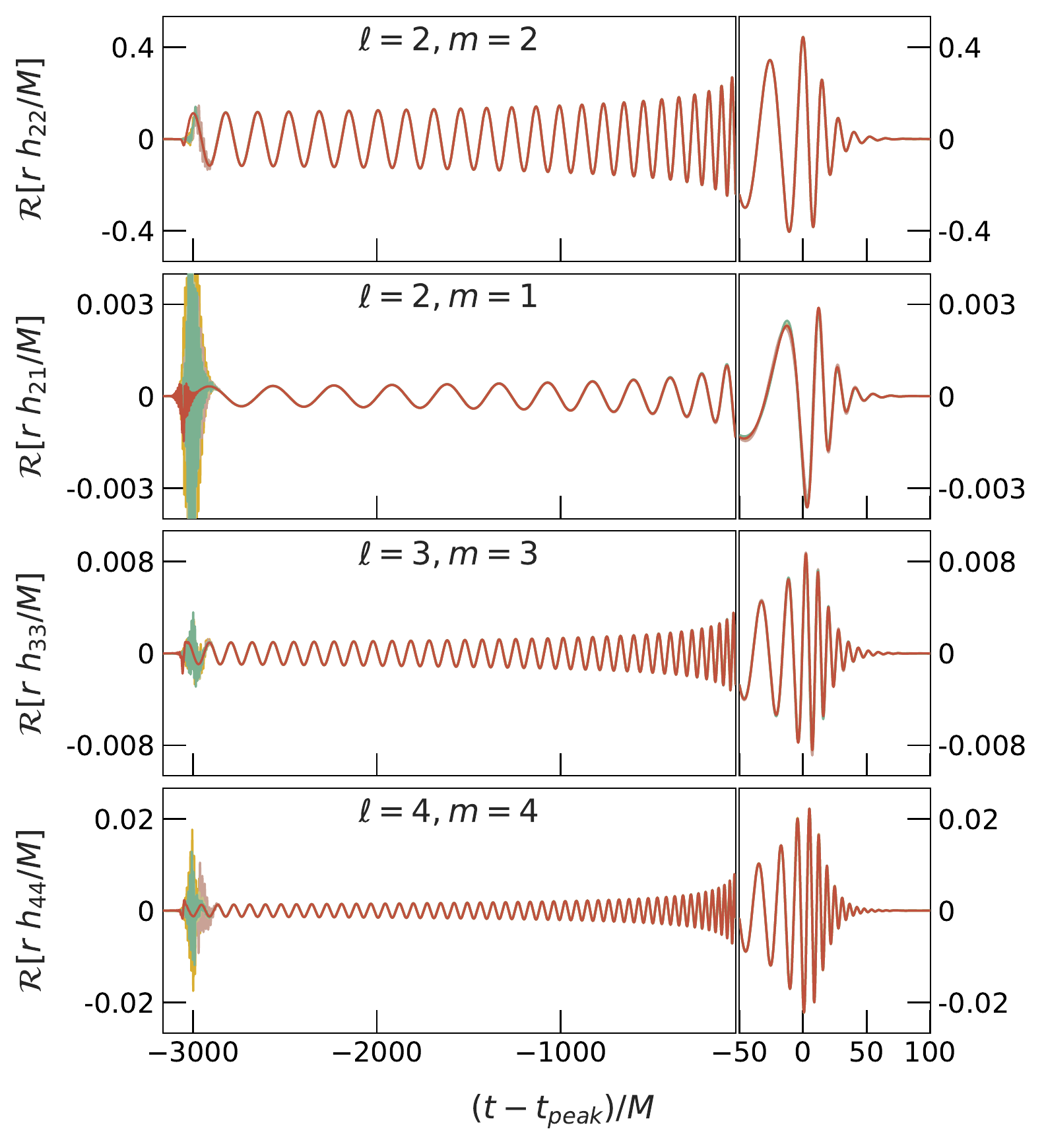}
\includegraphics[scale=0.63]{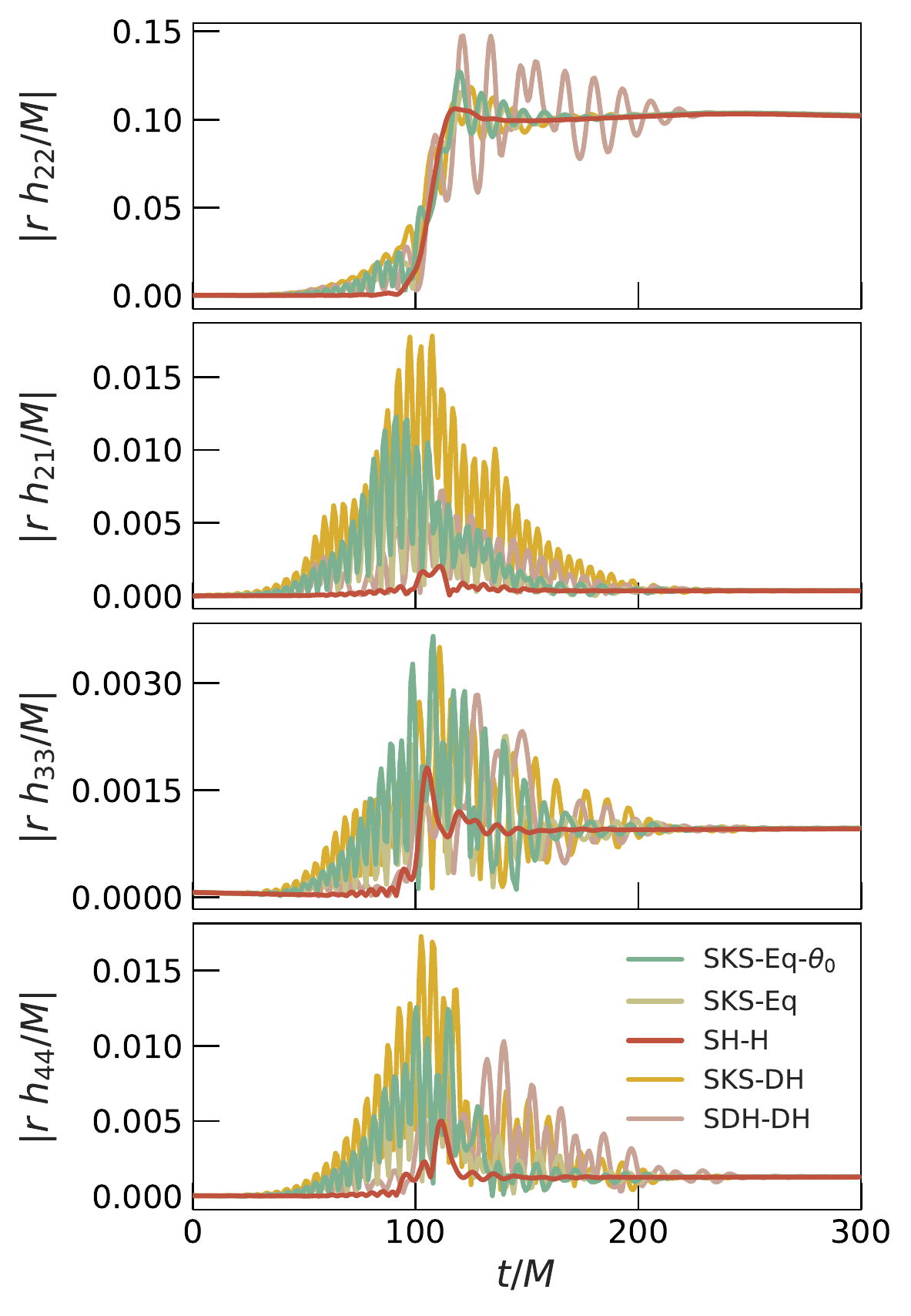}
\caption{Comparison of the waveforms resulting from evolution of different
initial data sets. The left column shows the real parts of different
spin-weighted spherical harmonic modes. The waveforms are aligned by
time-shifting them so that the peak amplitude occurs at $t=0$, and
phase-shifting them so that the orbital phase is zero at $t=0$. Once the junk
radiation leaves the domain, the waveforms agree very well between the
different initial data sets. The right panels show the amplitudes of
the different modes (without any time-shifting) during the junk
radiation stage. We see that \SHK initial data results in the least amount of
junk radiation. \SDH initial data, on the other hand, leads to the most junk
radiation. Note however, that junk radiation is not well resolved for all
cases except \SHK (cf. Fig~\ref{Fig:BBHEvConvTest}), hence the amount of junk
radiation changes significantly with resolution.
}
\label{Fig:Waveforms}
\end{center}
\end{figure*}

\begin{figure}[tbh]
\begin{center}
\includegraphics[scale=0.6]{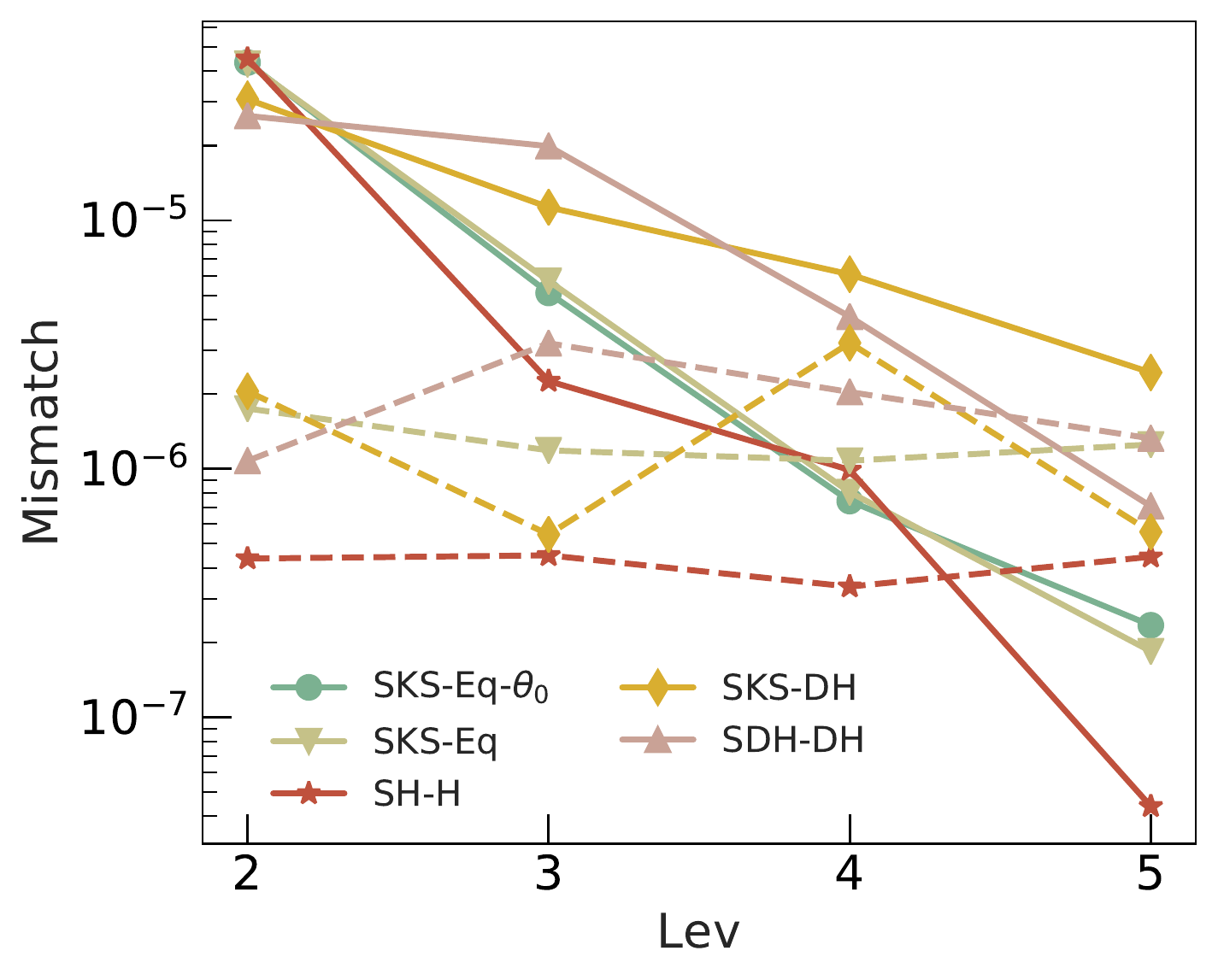}
\caption{
Median mismatches across the sky in the source frame
between waveforms generated from different initial data sets,
as a function of numerical resolution.
The horizontal axis shows the numerical resolution; we ran at five
different numerical resolutions labeled from lowest (Lev=1) to highest (Lev=5).
The solid lines represent numerical resolution error: they compare the
waveform at the labeled Lev to that of Lev$-1$.
Dashed lines show the differences between the waveform generated from evolving
the labeled initial data set to that generated from evolving \SKSHorBC.
The numerical resolution errors show reasonable convergence, as
expected. Interestingly, the mismatch between different initial data sets
does not change significantly with resolution. For sufficiently high
resolution, the resolution errors become smaller than initial data differences.
See discussion in Sec.~\ref{Subsec:WaveformComparison} for more details.
}
\label{Fig:IDvsResolutionMismatch}
\end{center}
\end{figure}

\subsection{Waveform comparison}
\label{Subsec:WaveformComparison}

Figure~\ref{Fig:Waveforms} shows the gravitational waveforms obtained
by the evolution of the different initial data sets. The waveforms are
extracted at different extraction radii up to $600M$ from the origin
and extrapolated to spatial infinity~\cite{Boyle-Mroue:2008}.
The left column shows different spin weighted
spherical harmonic modes of the waveform (we only show the real parts of the
modes here; the imaginary parts have very similar features).
As expected, after the initial junk radiation stage the waveforms between the
different initial data sets agree very well.

The right panels of Fig.~\ref{Fig:Waveforms} show the amplitudes of
different modes during the junk radiation stage. Among all the
initial data sets considered here, the junk radiation is the least in the
case of \SHK initial data. Compared to the current implementation in
SpEC (\SKSHorBC), the junk radiation decreases by a significant amount for \SHK
initial data. The junk radiation also leaves the system much faster in this
case.

As noted before, when evolving most initial data sets we perform a
smooth time-dependent gauge transformation so that the system settles
into damped harmonic gauge on a time scale of $50M$ after the start of
the evolution.  The \SDH and \SKSDH initial data sets already satisfy
the damped harmonic condition at $t=0$, so there is no need for such a
gauge transformation. We see that, among the cases considered, the junk
radiation is largest in the case of \SDH initial data. For \SKSDH initial data,
the junk radiation is at a comparable level to \SKSHorBC.
This suggests that we lose nothing by choosing the simpler
\SKSDH initial data over the standard choice of \SKSHorBC. We
also confirm that, as expected, the amount of junk radiation is roughly
independent of initial gauge, but depends on the free data.

We can quantify the agreement between any pair of waveforms by the
mismatch\footnote{We choose to use a flat noise curve so that our statements
are independent of the choice of GW detector.} between them:
\begin{gather}
\mathcal{MM} = 1 - \frac{\left< \hmath_1, \hmath_2 \right>}
{\sqrt{\left<\hmath_1,\hmath_1\right> \left<\hmath_2,\hmath_2\right>}}, \\
\left<\hmath_1, \hmath_2\right> = 4 \mathcal{R}
\left[ \int_{f_{min}}^{f_{max}}
\tilde{\hmath_1}(f) \tilde{\hmath_2^*}(f) ~df\right],
\end{gather}
where $\tilde{\hmath}_1(f)$ is the Fourier transform of $\hmath_1(t)$,
$\mathcal{R}[.]$ denotes the real part, $*$ denotes a complex conjugation,
and $f_{min}$ and $f_{max}$ denote the relevant frequency range.
$f_{min}$ is chosen to be the GW frequency at a time $500M$ from the start
(to exclude junk radiation) and $f_{max}$ is chosen to be $8$ times
the merger frequency of the $\ell=m=2$ mode.

We compute the mismatches as outlined in Appendix D of
Ref.~\cite{Blackman:2017dfb}, where both polarizations are treated on an equal
footing and the mismatch is minimized over shifts in time, initial binary
phase, and polarization angle. We
include
all available modes ($\ell \leq 8, |m| \leq \ell$), when computing the strain
\begin{gather}
\hmath(\theta, \phi, t)
    = \sum_{\ell,m} \,^{-2}Y_{\ell m}(\theta, \phi)~ h_{\ell m}(t),
\end{gather}
where $\,^{-2}Y_{lm}(\theta, \phi)$ are the spin-weighted spherical harmonics,
$\theta$ is the polar angle defined with respect to the initial orbital
angular momentum direction and $\phi$ is the azimuthal angle in the source
frame. We compute the mismatch for several different values of
$(\theta, \phi)$ (uniformly distributed in $\cos{\theta}$ and $\phi$) and
compare the median mismatches between different cases.

\begin{figure}[hbt]
\begin{center}
\includegraphics[scale=0.54]{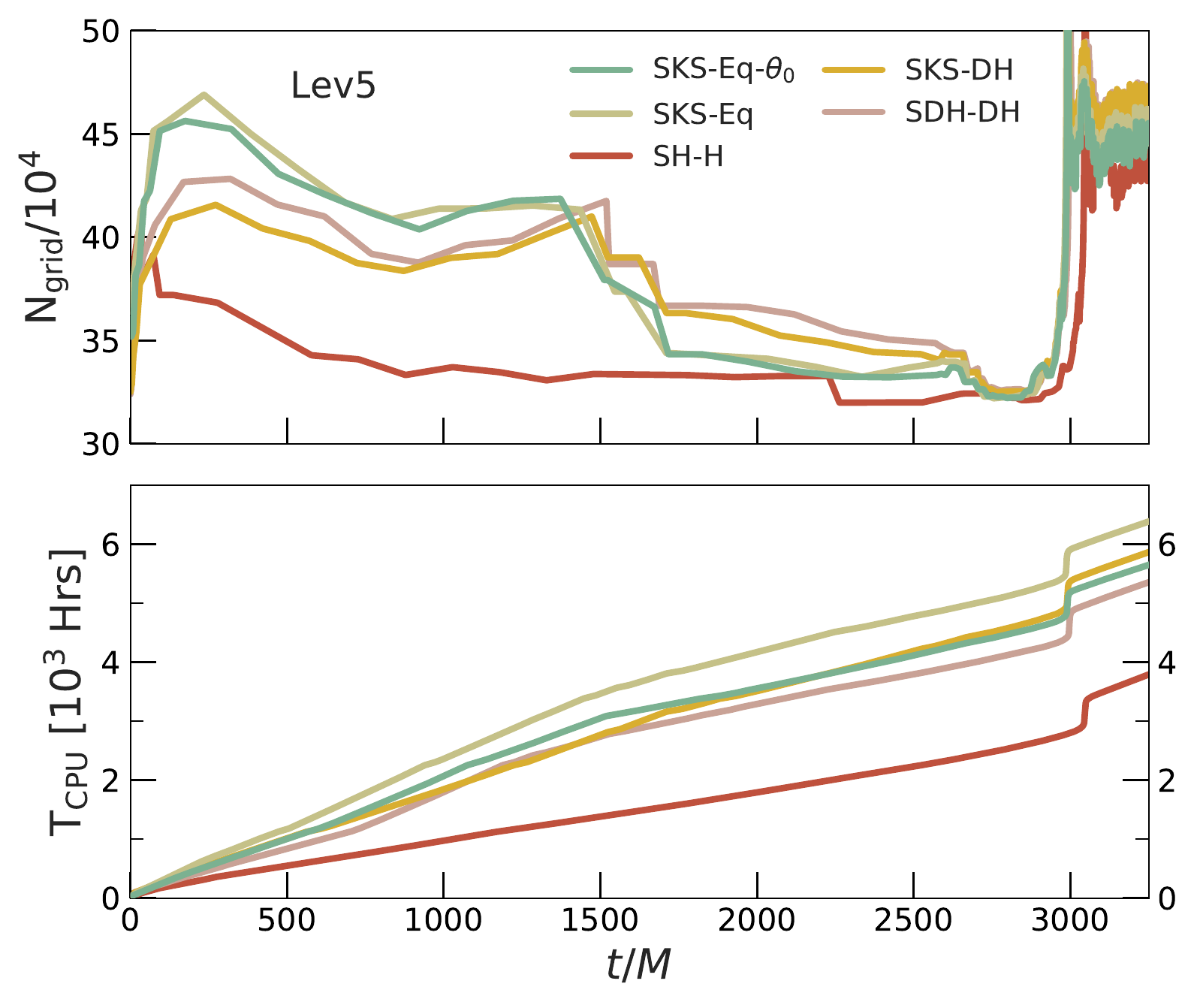}
\caption{Computational efficiency.
The top panel shows the total number of collocation points versus time for
several simulations running with the same AMR tolerance.
The bottom panel shows
the total CPU time as a function of the evolution
time. Using \SHK initial data speeds up the evolution by about $33\%$
compared to \SKSHorBC. All simulations are performed
on the same machine with the same number of CPUs.
}
\label{Fig:SimExpense}
\end{center}
\end{figure}

\begin{figure*}[hbt]
\begin{center}
\includegraphics[scale=0.475]{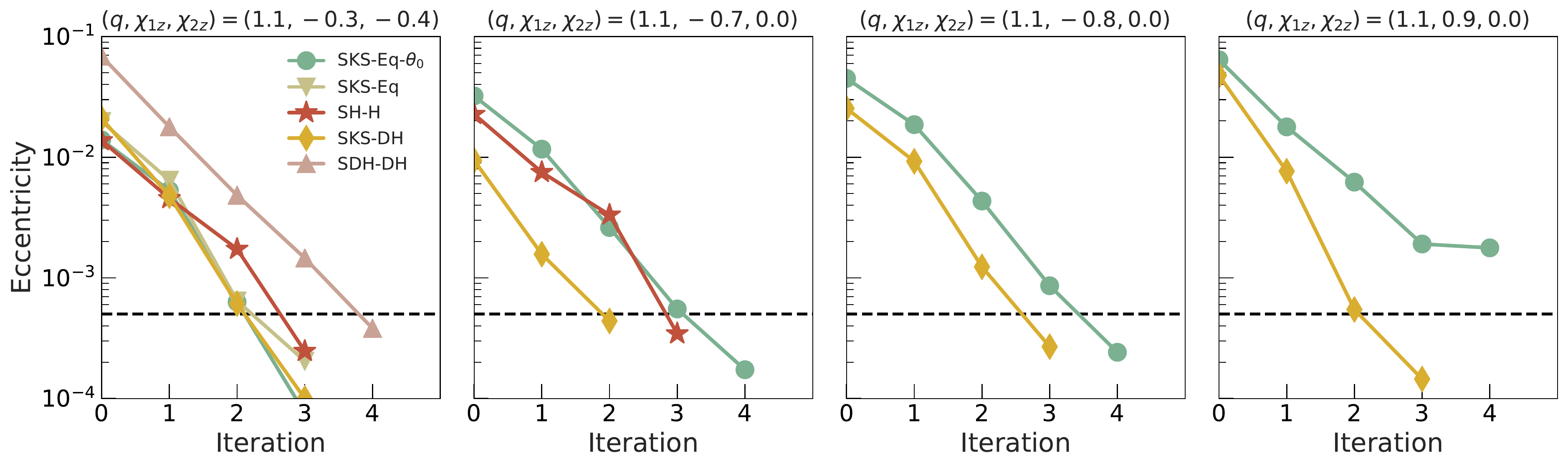}
\caption{Eccentricity reduction iterations for different initial data sets
considered in this study. The parameters of the binary are shown in the text
above each plot. We stop the iterations once the eccentricity reaches
$5\times10^{-4}$; this cutoff is shown as a black dashed line.
}
\label{Fig:EccRed}
\end{center}
\end{figure*}

Figure~\ref{Fig:IDvsResolutionMismatch} compares the median mismatches between
waveforms from different initial data sets to the median mismatch between
waveforms computed at different numerical resolutions. First, we note that the
numerical resolution errors show reasonable convergence, as expected.
Interestingly, we find that the differences between different initial data
sets does not change significantly with resolution. We understand this as
follows: Different initial data sets correspond to slightly different physical
systems (i.e. they have different junk radiation and therefore slightly
different orbital eccentricities and BH masses and spins, cf.
Fig.~\ref{Fig:CompParams} and Fig.~\ref{Fig:EccRed}) and this difference
is independent of resolution.  At low resolution, the differences due to
different initial data sets are within the numerical resolution errors, as was
found in Ref.~\cite{Garcia:2012dc}. However, contrary to the
findings\footnote{Note that Ref.~\cite{Garcia:2012dc} compares the phase and
amplitude of the quadrupole mode ($\ell=2$, $m=\pm2$) to evaluate the errors
between waveforms. Instead, we use the mismatch between the waveforms,
including all available modes, to evaluate the errors. Also,
Ref.~\cite{Garcia:2012dc} compares \SKSHorBC initial data to
CFMS (Conformally Flat Maximally Sliced) initial data, for an equal mass
non-spinning BBH.} of Ref.~\cite{Garcia:2012dc}, as we go towards high
resolution, the numerical resolution errors eventually go below the initial
data differences. This suggests that the resolution is now high enough to
differentiate between the initial data sets. These results also suggest that
when very high accuracy is required, one should be concerned with how well the
initial data set represents the desired astrophysical system.
Specifically, it is important to measure masses and spins after the
junk radiation, and one must consider tuning initial data parameters to
achieve desired ``post-junk'' parameters.

\subsection{Simulation expense}
\label{Subsec:SimExpense}

As discussed at the beginning of Sec.~\ref{Sec:BBhEv},
the resolution of a simulation is determined by specifying
an AMR error tolerance.  For different simulations, the same AMR tolerance may
result in a different number of collocation points and a different
computational expense, since AMR chooses the number of collocation points based
on the properties of the solution. Figure~\ref{Fig:SimExpense} shows the number
of collocation points in the domain (top panel) and the total CPU time (bottom
panel) for the different cases we consider, for a fixed AMR tolerance. For \SHK
initial data, not only is the constraint violation during the junk radiation
stage lower by an order of magnitude, this is achieved using $15\%$ fewer
collocation points and with a $33\%$ speed-up compared to \SKSHorBC.
This is another indication that evolutions of \SHK data contain fewer or
smaller high-frequency features than for other initial data sets, so that AMR
needs fewer collocation points to meet its error tolerance. These features can
possibly be physical high-frequency oscillations associated with junk
radiation, gauge oscillations, or gauge features that might manifest as
sharper features in quasi-stationary metric functions near the horizons.
We do not see significant differences in simulation
expense between \SKSHorBC and \SDH or \SKSDH initial data sets.
While this speedup is shown for the specific case of  $q=1.1$,
$\chi_{1z}=-0.3$, and $\chi_{2z}=-0.4$, we find similar improvements for more
generic cases as well.  However, since this improvement is largely due
to lesser junk content, we expect speed-ups only in the initial stages of the
evolution. For example, at times \footnote{The outer boundary for these
simulations is placed at a Euclidean radius of $800 M$, so $1600 M$ is
approximately the light crossing time for the domain, at which point the junk
radiation will have moved out of the domain.}
$t>1600M$ in Fig.~\ref{Fig:SimExpense},
the number of grid points and the CPU-time per simulation time are comparable
for \SHK and \SKSHorBC.  This also implies that the speed advantage of \SHK will
be less for longer simulations.

\subsection{Constructing zero-eccentricity initial data}
\label{Subsec:EccReduction}

Unlike the Newtonian or post-Newtonian (PN) case, in full general relativity
there is no analytic expression for the orbital parameters of two
compact objects that yield a zero-eccentricity orbit. In order to
achieve quasi-circular initial data, we adopt an iterative procedure as
follows~\cite{Buonanno:2010yk}: Start with an initial guess for orbital
parameters $\bold{\Omega_0}$ and $\dot{a}_0$ (defined in
Eq.~\ref{Eq:ShiftOuterBC}), typically taken from PN. Construct initial data
with these initial orbital parameters and evolve for $\sim2$ orbits,
compute the eccentricity from
the binary orbit and update the initial orbital parameters. Repeat until the
desired eccentricity is achieved.

Note that the eccentricity is measured over a few orbits of evolution, so that
updating the initial orbital parameters effectively involves an extrapolation
back in time to $t=0$. When there is also a gauge transformation happening
before or during the eccentricity measurement (cf.
Sec.~\ref{Subsubsec:dtLapseShift}), this extrapolation can in principle be
erroneous. Therefore, it is interesting to compare the eccentricity reduction
procedure for the different initial data sets we construct. Particularly for
\SKSDH and \SDH initial data sets, where there is no initial gauge
transformation, we might expect improvements in eccentricity reduction.
Figure~\ref{Fig:EccRed} shows the eccentricity reduction iterations for
different initial data sets. While we see that \SKSDH reaches the desired
eccentricity in fewer iterations than \SKSHorBC, we note that the initial guess
from PN theory produces lower starting eccentricity for this case. In general,
as the slopes of the curves do not differ significantly, we cannot conclusively
say that the eccentricity reduction procedure improves when there is no gauge
transformation. However, we find that \SKSDH is either better or the same as
\SKSHorBC for eccentricity reduction, for the cases we tested. Apart from \SDH
initial data, all other initial data sets seem to perform at the same level
as \SKSHorBC. For \SDH, while the rate of eccentricity reduction is the same,
the initial guess from post-Newtonian theory produces higher eccentricity
initial data. These results suggest that other approximations made in our
eccentricity-reduction procedure have a larger influence than
the effect of a time-dependent gauge transformation.

\section{Conclusion}
\label{Sec:Conclusion}

In this paper, we introduce new ways to choose free
data and new boundary conditions at excision surfaces, when
constructing BBH initial data. Furthermore, we experiment with several initial
gauge choices. We evolve these initial data sets and compare the
waveforms, junk radiation, evolution of component parameters, constraint
violations, simulation expense, and ease of constructing zero-eccentricity
initial data for the different cases.

The initial data cases we compare include the following new features
compared to the
traditional ``SKS'' initial data (here called \SKSHorBC) used in
past BBH simulations performed by the SpEC code:
\begin{itemize}
\item We introduce new boundary conditions
    that allow the initial-data numerical grid to extend
    inside (as opposed to on) the apparent horizons.
    Because the numerical grid for {\it evolution} must extend inside
    the apparent horizon, these new boundary conditions allow us
    to eliminate an extrapolation from the initial-data grid to the
    evolution grid.
    This reduces the initial constraint violations near
    the individual BH horizons by about 3 orders of magnitude. We denote the
    current implementation (\SKSHorBC) with only this change by \SKS.
  \item We construct BBH initial data with free data given by a
    superposition of two Harmonic-Kerr single BHs as derived in
    Ref.~\cite{cook_scheel97}. The initial gauge is imposed by setting
    $\partial_t N$ and $\partial_t N^i$ according to the harmonic
    gauge condition. We denote this by \SHK.
  \item We construct BBH initial data with free data given
    by a superposition of two
    Damped-Harmonic single BHs as derived in Ref.~\cite{Varma:2018sbhdh}. The
    initial gauge is imposed by setting $\partial_t N$ and $\partial_t N^i$
    according to the damped harmonic gauge condition. We denote this by \SDH.
  \item We also construct initial data identical to \SKS above, except
    $\partial_t N$ and $\partial_t N^i$ are chosen
    according to the damped harmonic gauge condition as opposed to
    a quasiequilibrium condition. We denote this by \SKSDH.
\end{itemize}
Note that among these cases, we use the negative expansion boundary
condition for all except \SKSHorBC and we do a gauge transformation
into the damped harmonic gauge over a time scale of $50M$ at the start
of evolution for all except \SDH and \SKSDH (which already satisfy this
gauge condition).

We compare these initial data sets by evolving a nonprecessing BBH system with
mass ratio $q=1.1$ and dimensionless spins $\chi_{1z}=-0.3$, $\chi_{2z}=-0.4$
along the orbital angular momentum direction. We compare the gravitational
waves (extrapolated to spatial infinity) generated using the different initial
data sets by computing the mismatches between them. We also compare these
mismatches to the mismatches between waveforms evolved at different numerical
resolution. As expected, the numerical resolution errors decrease as we go
towards higher resolutions. However, we find that the mismatches between
different  initial data sets are approximately independent of resolution; we
attribute this to the small physical differences between different initial
data sets. These differences correspond to different amounts of junk
radiation, and different parameters such as masses, spins, and orbital
eccentricity. At low resolution, the initial data differences are below the
numerical resolution errors. However, at high resolutions the numerical
truncation error eventually drops below the initial data differences.
Therefore, one must be careful to associate the waveform with the parameters
(masses, spins, orbital eccentricity) measured after the junk radiation stage
of the evolution rather than the parameters used to construct initial data.

\subsection{The case for using \SHK initial data}

\begin{figure}[hbt]
\begin{center}
\includegraphics[scale=0.24]{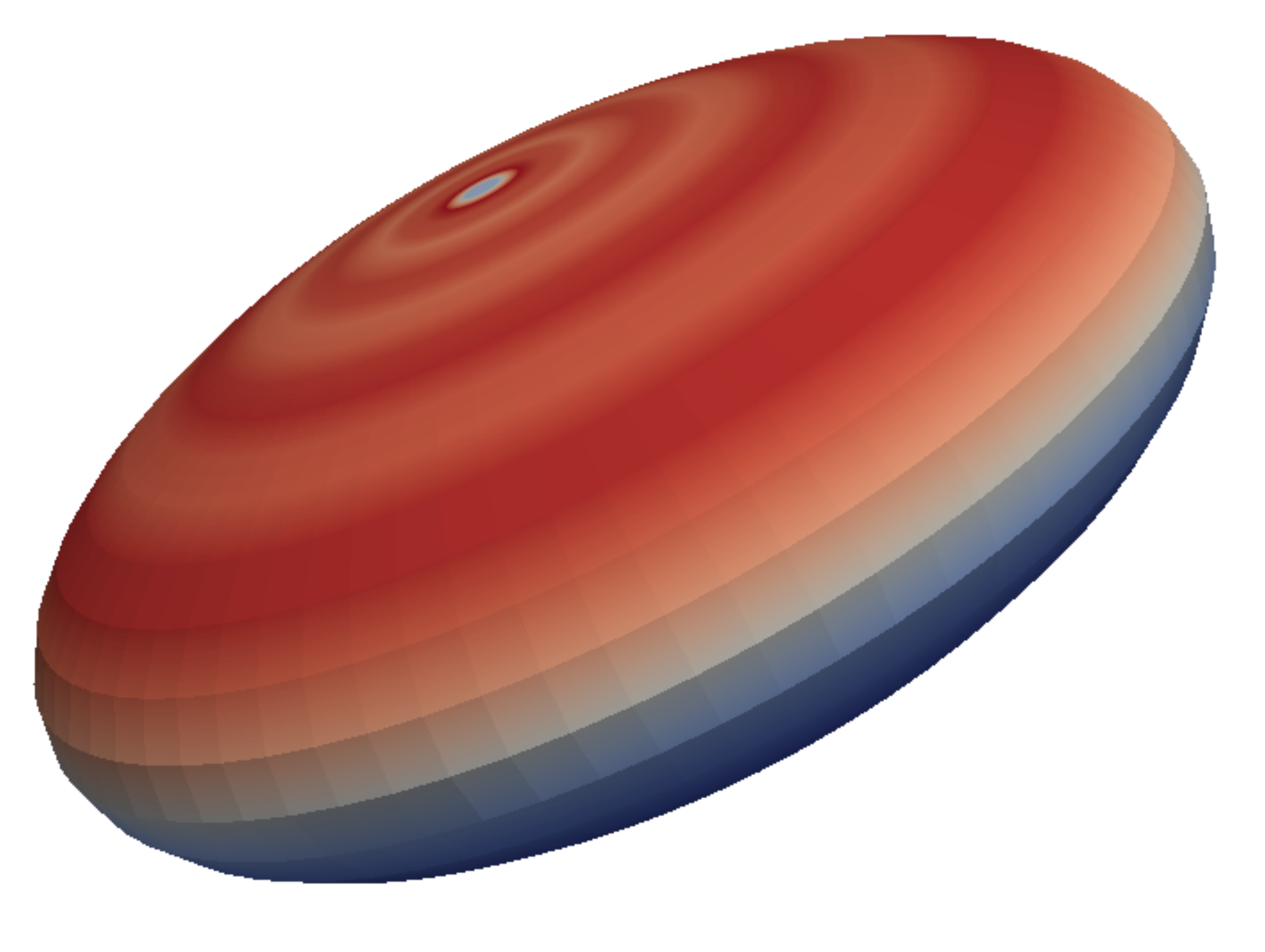}
~\includegraphics[scale=0.19]{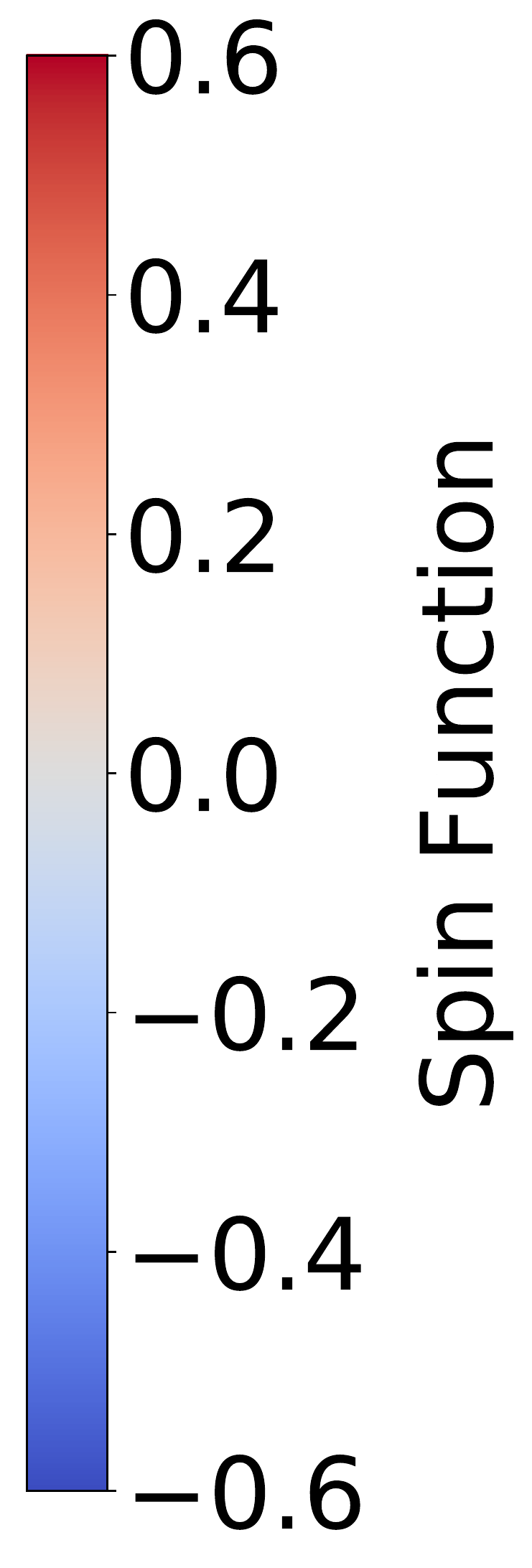}
\caption{Apparent horizon surface for a single BH with dimensionless spin
$\chi=0.9$ in the harmonic coordinates of Ref.~\cite{cook_scheel97}.
The colors show the imaginary part of complex scalar curvature of the 2D
horizon surface~\cite{Owen2009, OwenEtAl:2011}. The spin direction is along
the poles. We note that the shape of the surface is compressed in the spin
direction (much like a pancake), making it difficult to construct initial data.
The ratio of the extents of the horizon between the spin direction and an
orthogonal direction goes as $\sqrt{1-\chi^2}$, so this issue becomes more
prominent at high spins. We currently can construct superposed harmonic
initial data only for spins $\chi\leq0.7$.
}
\label{Fig:Pancake}
\end{center}
\end{figure}

By comparing the different initial data sets we conclude that \SHK initial data
has the following benefits over the current implementation in SpEC (\SKSHorBC):
\begin{itemize}
\setlength\itemsep{1em}
\item The initial spurious junk radiation is much smaller.
\item The junk radiation leaves the system sooner.
\item The constraint violations during the junk radiation stage decrease by
    about an order of magnitude.
\item The constraints have good convergence even during junk radiation. This
    suggests that the junk radiation is being resolved properly.
\item The time variation in masses and spins during junk radiation
    is smaller by an order of magnitude.
\item This improvement in constraints during junk radiation is achieved using
$15\%$ less collocation points in the domain. This leads to a remarkable
$33\%$ speed up in the total evolution time.
\end{itemize}

Because of these benefits, we recommend \SHK as the preferred choice
for initial data, when possible. Unfortunately,
we are currently able to construct \SHK initial data only for dimensionless
spin magnitudes $\chi \leq 0.7$. At higher spins the single BH harmonic
coordinates used for the construction of the free data in XCTS are too
distorted (see Fig.~\ref{Fig:Pancake}), and the elliptic solver fails to
converge. Therefore, we recommend that \SHK initial data be used for
$\chi \leq 0.7$; otherwise, \SKSDH is our recommendation,
since \SKSDH eliminates the need for extrapolation and for
dynamical gauge changes, and it performs no worse than \SKSHorBC.

\subsection{Outlook and future work}
Having seen that \SHK initial data is superior to the current implementation
in SpEC, it would be worthwhile to extend it to spins higher than $\chi=0.7$.
To overcome the problem with highly distorted horizons, one could use a
coordinate map to make the horizons more spherical; this may violate
the harmonic spatial gauge condition but will preserve harmonic time slicing.
It would be interesting to
see if such a map preserves the benefits of \SHK initial data.

Our tests on \SHK initial data suggest that even the junk radiation
stage is convergent when using this initial data.  Therefore, \SHK
initial data allows us to study properties of junk radiation
transients, such as their frequency content or how long they remain in
the computational domain.  For other initial data sets, the main
obstacle for such a study is the prohibitively high resolution needed
to fully resolve junk radiation.

\begin{acknowledgments}
We thank Geoffrey Lovelace, Saul Teukolsky and Leo Stein for useful
discussions. This work was supported in part by the Sherman Fairchild
Foundation and NSF grants PHY-1404569, PHY-170212, and PHY-1708213 at Caltech.
The simulations were performed on the Wheeler cluster at Caltech,
which is supported by the Sherman Fairchild Foundation and Caltech.

\end{acknowledgments}

\bibliography{References}

\begin{thebibliography}{60}%
\makeatletter
\providecommand \@ifxundefined [1]{%
 \@ifx{#1\undefined}
}%
\providecommand \@ifnum [1]{%
 \ifnum #1\expandafter \@firstoftwo
 \else \expandafter \@secondoftwo
 \fi
}%
\providecommand \@ifx [1]{%
 \ifx #1\expandafter \@firstoftwo
 \else \expandafter \@secondoftwo
 \fi
}%
\providecommand \natexlab [1]{#1}%
\providecommand \enquote  [1]{``#1''}%
\providecommand \bibnamefont  [1]{#1}%
\providecommand \bibfnamefont [1]{#1}%
\providecommand \citenamefont [1]{#1}%
\providecommand \href@noop [0]{\@secondoftwo}%
\providecommand \href [0]{\begingroup \@sanitize@url \@href}%
\providecommand \@href[1]{\@@startlink{#1}\@@href}%
\providecommand \@@href[1]{\endgroup#1\@@endlink}%
\providecommand \@sanitize@url [0]{\catcode `\\12\catcode `\$12\catcode
  `\&12\catcode `\#12\catcode `\^12\catcode `\_12\catcode `\%12\relax}%
\providecommand \@@startlink[1]{}%
\providecommand \@@endlink[0]{}%
\providecommand \url  [0]{\begingroup\@sanitize@url \@url }%
\providecommand \@url [1]{\endgroup\@href {#1}{\urlprefix }}%
\providecommand \urlprefix  [0]{URL }%
\providecommand \Eprint [0]{\href }%
\providecommand \doibase [0]{http://dx.doi.org/}%
\providecommand \selectlanguage [0]{\@gobble}%
\providecommand \bibinfo  [0]{\@secondoftwo}%
\providecommand \bibfield  [0]{\@secondoftwo}%
\providecommand \translation [1]{[#1]}%
\providecommand \BibitemOpen [0]{}%
\providecommand \bibitemStop [0]{}%
\providecommand \bibitemNoStop [0]{.\EOS\space}%
\providecommand \EOS [0]{\spacefactor3000\relax}%
\providecommand \BibitemShut  [1]{\csname bibitem#1\endcsname}%
\let\auto@bib@innerbib\@empty
\bibitem [{\citenamefont {Khan}\ \emph {et~al.}(2016)\citenamefont {Khan},
  \citenamefont {Husa}, \citenamefont {Hannam}, \citenamefont {Ohme},
  \citenamefont {Pürrer}, \citenamefont {Jiménez~Forteza},\ and\
  \citenamefont {Bohé}}]{Khan:2015jqa}%
  \BibitemOpen
  \bibfield  {author} {\bibinfo {author} {\bibfnamefont {S.}~\bibnamefont
  {Khan}}, \bibinfo {author} {\bibfnamefont {S.}~\bibnamefont {Husa}}, \bibinfo
  {author} {\bibfnamefont {M.}~\bibnamefont {Hannam}}, \bibinfo {author}
  {\bibfnamefont {F.}~\bibnamefont {Ohme}}, \bibinfo {author} {\bibfnamefont
  {M.}~\bibnamefont {Pürrer}}, \bibinfo {author} {\bibfnamefont
  {X.}~\bibnamefont {Jiménez~Forteza}}, \ and\ \bibinfo {author}
  {\bibfnamefont {A.}~\bibnamefont {Bohé}},\ }\href {\doibase
  10.1103/PhysRevD.93.044007} {\bibfield  {journal} {\bibinfo  {journal} {Phys.
  Rev.}\ }\textbf {\bibinfo {volume} {D93}},\ \bibinfo {pages} {044007}
  (\bibinfo {year} {2016})},\ \Eprint {http://arxiv.org/abs/1508.07253}
  {arXiv:1508.07253 [gr-qc]} \BibitemShut {NoStop}%
\bibitem [{\citenamefont {Hannam}\ \emph {et~al.}(2014)\citenamefont {Hannam},
  \citenamefont {Schmidt}, \citenamefont {Boh{\' e}}, \citenamefont {Haegel},
  \citenamefont {Husa} \emph {et~al.}}]{Hannam:2013oca}%
  \BibitemOpen
  \bibfield  {author} {\bibinfo {author} {\bibfnamefont {M.}~\bibnamefont
  {Hannam}}, \bibinfo {author} {\bibfnamefont {P.}~\bibnamefont {Schmidt}},
  \bibinfo {author} {\bibfnamefont {A.}~\bibnamefont {Boh{\' e}}}, \bibinfo
  {author} {\bibfnamefont {L.}~\bibnamefont {Haegel}}, \bibinfo {author}
  {\bibfnamefont {S.}~\bibnamefont {Husa}},  \emph {et~al.},\ }\href {\doibase
  10.1103/PhysRevLett.113.151101} {\bibfield  {journal} {\bibinfo  {journal}
  {Phys.\ Rev.\ Lett.}\ }\textbf {\bibinfo {volume} {113}},\ \bibinfo {pages}
  {151101} (\bibinfo {year} {2014})},\ \Eprint {http://arxiv.org/abs/1308.3271}
  {arXiv:1308.3271 [gr-qc]} \BibitemShut {NoStop}%
\bibitem [{\citenamefont {Boh\'e}\ \emph {et~al.}(2017)\citenamefont {Boh\'e},
  \citenamefont {Shao}, \citenamefont {Taracchini}, \citenamefont {Buonanno},
  \citenamefont {Babak}, \citenamefont {Harry}, \citenamefont {Hinder},
  \citenamefont {Ossokine}, \citenamefont {P\"urrer}, \citenamefont {Raymond},
  \citenamefont {Chu}, \citenamefont {Fong}, \citenamefont {Kumar},
  \citenamefont {Pfeiffer}, \citenamefont {Boyle}, \citenamefont {Hemberger},
  \citenamefont {Kidder}, \citenamefont {Lovelace}, \citenamefont {Scheel},\
  and\ \citenamefont {Szil\'agyi}}]{Bohe:2016gbl}%
  \BibitemOpen
  \bibfield  {author} {\bibinfo {author} {\bibfnamefont {A.}~\bibnamefont
  {Boh\'e}}, \bibinfo {author} {\bibfnamefont {L.}~\bibnamefont {Shao}},
  \bibinfo {author} {\bibfnamefont {A.}~\bibnamefont {Taracchini}}, \bibinfo
  {author} {\bibfnamefont {A.}~\bibnamefont {Buonanno}}, \bibinfo {author}
  {\bibfnamefont {S.}~\bibnamefont {Babak}}, \bibinfo {author} {\bibfnamefont
  {I.~W.}\ \bibnamefont {Harry}}, \bibinfo {author} {\bibfnamefont
  {I.}~\bibnamefont {Hinder}}, \bibinfo {author} {\bibfnamefont
  {S.}~\bibnamefont {Ossokine}}, \bibinfo {author} {\bibfnamefont
  {M.}~\bibnamefont {P\"urrer}}, \bibinfo {author} {\bibfnamefont
  {V.}~\bibnamefont {Raymond}}, \bibinfo {author} {\bibfnamefont
  {T.}~\bibnamefont {Chu}}, \bibinfo {author} {\bibfnamefont {H.}~\bibnamefont
  {Fong}}, \bibinfo {author} {\bibfnamefont {P.}~\bibnamefont {Kumar}},
  \bibinfo {author} {\bibfnamefont {H.~P.}\ \bibnamefont {Pfeiffer}}, \bibinfo
  {author} {\bibfnamefont {M.}~\bibnamefont {Boyle}}, \bibinfo {author}
  {\bibfnamefont {D.~A.}\ \bibnamefont {Hemberger}}, \bibinfo {author}
  {\bibfnamefont {L.~E.}\ \bibnamefont {Kidder}}, \bibinfo {author}
  {\bibfnamefont {G.}~\bibnamefont {Lovelace}}, \bibinfo {author}
  {\bibfnamefont {M.~A.}\ \bibnamefont {Scheel}}, \ and\ \bibinfo {author}
  {\bibfnamefont {B.}~\bibnamefont {Szil\'agyi}},\ }\href {\doibase
  10.1103/PhysRevD.95.044028} {\bibfield  {journal} {\bibinfo  {journal} {Phys.
  Rev. D}\ }\textbf {\bibinfo {volume} {95}},\ \bibinfo {pages} {044028}
  (\bibinfo {year} {2017})},\ \Eprint {http://arxiv.org/abs/1611.03703}
  {arXiv:1611.03703 [gr-qc]} \BibitemShut {NoStop}%
\bibitem [{\citenamefont {{Pan}}\ \emph {et~al.}(2013)\citenamefont {{Pan}},
  \citenamefont {{Buonanno}}, \citenamefont {{Taracchini}}, \citenamefont
  {{Kidder}}, \citenamefont {{Mrou{\'e}}}, \citenamefont {{Pfeiffer}},
  \citenamefont {{Scheel}},\ and\ \citenamefont
  {{Szil{\'a}gyi}}}]{Pan:2013rra}%
  \BibitemOpen
  \bibfield  {author} {\bibinfo {author} {\bibfnamefont {Y.}~\bibnamefont
  {{Pan}}}, \bibinfo {author} {\bibfnamefont {A.}~\bibnamefont {{Buonanno}}},
  \bibinfo {author} {\bibfnamefont {A.}~\bibnamefont {{Taracchini}}}, \bibinfo
  {author} {\bibfnamefont {L.~E.}\ \bibnamefont {{Kidder}}}, \bibinfo {author}
  {\bibfnamefont {A.~H.}\ \bibnamefont {{Mrou{\'e}}}}, \bibinfo {author}
  {\bibfnamefont {H.~P.}\ \bibnamefont {{Pfeiffer}}}, \bibinfo {author}
  {\bibfnamefont {M.~A.}\ \bibnamefont {{Scheel}}}, \ and\ \bibinfo {author}
  {\bibfnamefont {B.}~\bibnamefont {{Szil{\'a}gyi}}},\ }\href@noop {}
  {\bibfield  {journal} {\bibinfo  {journal} {Phys.\ Rev.\ D}\ }\textbf
  {\bibinfo {volume} {89}},\ \bibinfo {pages} {084006} (\bibinfo {year}
  {2013})},\ \Eprint {http://arxiv.org/abs/1307.6232} {arXiv:1307.6232 [gr-qc]}
  \BibitemShut {NoStop}%
\bibitem [{\citenamefont {Taracchini}\ \emph {et~al.}(2014)\citenamefont
  {Taracchini}, \citenamefont {Buonanno}, \citenamefont {Pan}, \citenamefont
  {Hinderer}, \citenamefont {Boyle}, \citenamefont {Hemberger}, \citenamefont
  {Kidder}, \citenamefont {Lovelace}, \citenamefont {Mroue}, \citenamefont
  {Pfeiffer}, \citenamefont {Scheel}, \citenamefont {Szil{\'a}gyi},
  \citenamefont {Taylor},\ and\ \citenamefont
  {Zenginoglu}}]{Taracchini:2013rva}%
  \BibitemOpen
  \bibfield  {author} {\bibinfo {author} {\bibfnamefont {A.}~\bibnamefont
  {Taracchini}}, \bibinfo {author} {\bibfnamefont {A.}~\bibnamefont
  {Buonanno}}, \bibinfo {author} {\bibfnamefont {Y.}~\bibnamefont {Pan}},
  \bibinfo {author} {\bibfnamefont {T.}~\bibnamefont {Hinderer}}, \bibinfo
  {author} {\bibfnamefont {M.}~\bibnamefont {Boyle}}, \bibinfo {author}
  {\bibfnamefont {D.~A.}\ \bibnamefont {Hemberger}}, \bibinfo {author}
  {\bibfnamefont {L.~E.}\ \bibnamefont {Kidder}}, \bibinfo {author}
  {\bibfnamefont {G.}~\bibnamefont {Lovelace}}, \bibinfo {author}
  {\bibfnamefont {A.~H.}\ \bibnamefont {Mroue}}, \bibinfo {author}
  {\bibfnamefont {H.~P.}\ \bibnamefont {Pfeiffer}}, \bibinfo {author}
  {\bibfnamefont {M.~A.}\ \bibnamefont {Scheel}}, \bibinfo {author}
  {\bibfnamefont {B.}~\bibnamefont {Szil{\'a}gyi}}, \bibinfo {author}
  {\bibfnamefont {N.~W.}\ \bibnamefont {Taylor}}, \ and\ \bibinfo {author}
  {\bibfnamefont {A.}~\bibnamefont {Zenginoglu}},\ }\href@noop {} {\bibfield
  {journal} {\bibinfo  {journal} {Phys.\ Rev.\ D}\ }\textbf {\bibinfo {volume}
  {89 (R)}},\ \bibinfo {pages} {061502} (\bibinfo {year} {2014})},\ \Eprint
  {http://arxiv.org/abs/1311.2544} {arXiv:1311.2544 [gr-qc]} \BibitemShut
  {NoStop}%
\bibitem [{\citenamefont {Abbott}\ \emph
  {et~al.}(2016{\natexlab{a}})\citenamefont {Abbott} \emph
  {et~al.}}]{LIGOVirgo2016a}%
  \BibitemOpen
  \bibfield  {author} {\bibinfo {author} {\bibfnamefont {B.~P.}\ \bibnamefont
  {Abbott}} \emph {et~al.} (\bibinfo {collaboration} {LIGO Scientific
  Collaboration, Virgo Collaboration}),\ }\href {\doibase
  10.1103/PhysRevLett.116.061102} {\bibfield  {journal} {\bibinfo  {journal}
  {Phys.\ Rev.\ Lett.}\ }\textbf {\bibinfo {volume} {116}},\ \bibinfo {pages}
  {061102} (\bibinfo {year} {2016}{\natexlab{a}})},\ \Eprint
  {http://arxiv.org/abs/1602.03837} {arXiv:1602.03837 [gr-qc]} \BibitemShut
  {NoStop}%
\bibitem [{\citenamefont {Abbott}\ \emph
  {et~al.}(2016{\natexlab{b}})\citenamefont {Abbott} \emph
  {et~al.}}]{Abbott:2016nmj}%
  \BibitemOpen
  \bibfield  {author} {\bibinfo {author} {\bibfnamefont {B.~P.}\ \bibnamefont
  {Abbott}} \emph {et~al.} (\bibinfo {collaboration} {LIGO Scientific
  Collaboration, Virgo Collaboration}),\ }\href {\doibase
  10.1103/PhysRevLett.116.241103} {\bibfield  {journal} {\bibinfo  {journal}
  {Phys. Rev. Lett.}\ }\textbf {\bibinfo {volume} {116}},\ \bibinfo {pages}
  {241103} (\bibinfo {year} {2016}{\natexlab{b}})},\ \Eprint
  {http://arxiv.org/abs/1606.04855} {arXiv:1606.04855 [gr-qc]} \BibitemShut
  {NoStop}%
\bibitem [{\citenamefont {Abbott}\ \emph
  {et~al.}(2017{\natexlab{a}})\citenamefont {Abbott} \emph
  {et~al.}}]{Abbott:2017gyy}%
  \BibitemOpen
  \bibfield  {author} {\bibinfo {author} {\bibfnamefont {B.~P.}\ \bibnamefont
  {Abbott}} \emph {et~al.} (\bibinfo {collaboration} {Virgo, LIGO
  Scientific}),\ }\href {\doibase 10.3847/2041-8213/aa9f0c} {\bibfield
  {journal} {\bibinfo  {journal} {Astrophys. J.}\ }\textbf {\bibinfo {volume}
  {851}},\ \bibinfo {pages} {L35} (\bibinfo {year} {2017}{\natexlab{a}})},\
  \Eprint {http://arxiv.org/abs/1711.05578} {arXiv:1711.05578 [astro-ph.HE]}
  \BibitemShut {NoStop}%
\bibitem [{\citenamefont {Abbott}\ \emph
  {et~al.}(2017{\natexlab{b}})\citenamefont {Abbott} \emph
  {et~al.}}]{Abbott:2017oio}%
  \BibitemOpen
  \bibfield  {author} {\bibinfo {author} {\bibfnamefont {B.~P.}\ \bibnamefont
  {Abbott}} \emph {et~al.} (\bibinfo {collaboration} {Virgo, LIGO
  Scientific}),\ }\href {\doibase 10.1103/PhysRevLett.119.141101} {\bibfield
  {journal} {\bibinfo  {journal} {Phys. Rev. Lett.}\ }\textbf {\bibinfo
  {volume} {119}},\ \bibinfo {pages} {141101} (\bibinfo {year}
  {2017}{\natexlab{b}})},\ \Eprint {http://arxiv.org/abs/1709.09660}
  {arXiv:1709.09660 [gr-qc]} \BibitemShut {NoStop}%
\bibitem [{\citenamefont {Abbott}\ \emph
  {et~al.}(2017{\natexlab{c}})\citenamefont {Abbott} \emph
  {et~al.}}]{Abbott:2017vtc}%
  \BibitemOpen
  \bibfield  {author} {\bibinfo {author} {\bibfnamefont {B.~P.}\ \bibnamefont
  {Abbott}} \emph {et~al.} (\bibinfo {collaboration} {VIRGO, LIGO
  Scientific}),\ }\href {\doibase 10.1103/PhysRevLett.118.221101} {\bibfield
  {journal} {\bibinfo  {journal} {Phys. Rev. Lett.}\ }\textbf {\bibinfo
  {volume} {118}},\ \bibinfo {pages} {221101} (\bibinfo {year}
  {2017}{\natexlab{c}})},\ \Eprint {http://arxiv.org/abs/1706.01812}
  {arXiv:1706.01812 [gr-qc]} \BibitemShut {NoStop}%
\bibitem [{\citenamefont {Aasi}\ \emph {et~al.}(2015)\citenamefont {Aasi} \emph
  {et~al.}}]{aLIGO2}%
  \BibitemOpen
  \bibfield  {author} {\bibinfo {author} {\bibfnamefont {J.}~\bibnamefont
  {Aasi}} \emph {et~al.} (\bibinfo {collaboration} {LIGO Scientific
  Collaboration}),\ }\href {\doibase 10.1088/0264-9381/32/7/074001} {\bibfield
  {journal} {\bibinfo  {journal} {Class.\ Quantum Grav.}\ }\textbf {\bibinfo
  {volume} {32}},\ \bibinfo {pages} {074001} (\bibinfo {year} {2015})},\
  \Eprint {http://arxiv.org/abs/1411.4547} {arXiv:1411.4547 [gr-qc]}
  \BibitemShut {NoStop}%
\bibitem [{\citenamefont {Abbott}\ \emph
  {et~al.}(2016{\natexlab{c}})\citenamefont {Abbott} \emph
  {et~al.}}]{TheLIGOScientific:2016wfe}%
  \BibitemOpen
  \bibfield  {author} {\bibinfo {author} {\bibfnamefont {B.~P.}\ \bibnamefont
  {Abbott}} \emph {et~al.} (\bibinfo {collaboration} {LIGO Scientific
  Collaboration, Virgo Collaboration}),\ }\href {\doibase
  10.1103/PhysRevLett.116.241102} {\bibfield  {journal} {\bibinfo  {journal}
  {Phys.~Rev.~Lett.}\ }\textbf {\bibinfo {volume} {116}},\ \bibinfo {pages}
  {241102} (\bibinfo {year} {2016}{\natexlab{c}})},\ \Eprint
  {http://arxiv.org/abs/1602.03840} {arXiv:1602.03840 [gr-qc]} \BibitemShut
  {NoStop}%
\bibitem [{\citenamefont {Abbott}\ \emph
  {et~al.}(2016{\natexlab{d}})\citenamefont {Abbott} \emph
  {et~al.}}]{TheLIGOScientific:2016src}%
  \BibitemOpen
  \bibfield  {author} {\bibinfo {author} {\bibfnamefont {B.~P.}\ \bibnamefont
  {Abbott}} \emph {et~al.} (\bibinfo {collaboration} {LIGO Scientific
  Collaboration, Virgo Collaboration}),\ }\href@noop {} {\bibfield  {journal}
  {\bibinfo  {journal} {Phys.~Rev.~Lett.}\ }\textbf {\bibinfo {volume} {116}},\
  \bibinfo {pages} {221101} (\bibinfo {year} {2016}{\natexlab{d}})},\ \Eprint
  {http://arxiv.org/abs/1602.03841} {arXiv:1602.03841 [gr-qc]} \BibitemShut
  {NoStop}%
\bibitem [{\citenamefont {{Garcia}}\ \emph {et~al.}(2012)\citenamefont
  {{Garcia}}, \citenamefont {{Lovelace}}, \citenamefont {{Kidder}},
  \citenamefont {{Boyle}}, \citenamefont {{Teukolsky}}, \citenamefont
  {{Scheel}},\ and\ \citenamefont {{Szil{\'a}gyi}}}]{Garcia:2012dc}%
  \BibitemOpen
  \bibfield  {author} {\bibinfo {author} {\bibfnamefont {B.}~\bibnamefont
  {{Garcia}}}, \bibinfo {author} {\bibfnamefont {G.}~\bibnamefont
  {{Lovelace}}}, \bibinfo {author} {\bibfnamefont {L.~E.}\ \bibnamefont
  {{Kidder}}}, \bibinfo {author} {\bibfnamefont {M.}~\bibnamefont {{Boyle}}},
  \bibinfo {author} {\bibfnamefont {S.~A.}\ \bibnamefont {{Teukolsky}}},
  \bibinfo {author} {\bibfnamefont {M.~A.}\ \bibnamefont {{Scheel}}}, \ and\
  \bibinfo {author} {\bibfnamefont {B.}~\bibnamefont {{Szil{\'a}gyi}}},\ }\href
  {\doibase 10.1103/PhysRevD.86.084054} {\bibfield  {journal} {\bibinfo
  {journal} {Phys.\ Rev.\ D}\ }\textbf {\bibinfo {volume} {86}},\ \bibinfo
  {eid} {084054} (\bibinfo {year} {2012})},\ \Eprint
  {http://arxiv.org/abs/1206.2943} {arXiv:1206.2943 [gr-qc]} \BibitemShut
  {NoStop}%
\bibitem [{SpE()}]{SpECWebsite}%
  \BibitemOpen
  \href@noop {} {}\bibinfo {howpublished}
  {\url{http://www.black-holes.org/SpEC.html}}\BibitemShut {NoStop}%
\bibitem [{\citenamefont {York}(1999)}]{York1999}%
  \BibitemOpen
  \bibfield  {author} {\bibinfo {author} {\bibfnamefont {J.~W.}\ \bibnamefont
  {York}},\ }\href {\doibase 10.1103/PhysRevLett.82.1350} {\bibfield  {journal}
  {\bibinfo  {journal} {Phys.\ Rev.\ Lett.}\ }\textbf {\bibinfo {volume}
  {82}},\ \bibinfo {pages} {1350} (\bibinfo {year} {1999})}\BibitemShut
  {NoStop}%
\bibitem [{\citenamefont {Pfeiffer}\ and\ \citenamefont
  {York}(2003)}]{Pfeiffer2003b}%
  \BibitemOpen
  \bibfield  {author} {\bibinfo {author} {\bibfnamefont {H.~P.}\ \bibnamefont
  {Pfeiffer}}\ and\ \bibinfo {author} {\bibfnamefont {J.~W.}\ \bibnamefont
  {York}},\ }\href {\doibase 10.1103/PhysRevD.67.044022} {\bibfield  {journal}
  {\bibinfo  {journal} {Phys.\ Rev.\ D}\ }\textbf {\bibinfo {volume} {67}},\
  \bibinfo {pages} {044022} (\bibinfo {year} {2003})}\BibitemShut {NoStop}%
\bibitem [{\citenamefont {Lovelace}\ \emph {et~al.}(2008)\citenamefont
  {Lovelace}, \citenamefont {Owen}, \citenamefont {Pfeiffer},\ and\
  \citenamefont {Chu}}]{Lovelace2008}%
  \BibitemOpen
  \bibfield  {author} {\bibinfo {author} {\bibfnamefont {G.}~\bibnamefont
  {Lovelace}}, \bibinfo {author} {\bibfnamefont {R.}~\bibnamefont {Owen}},
  \bibinfo {author} {\bibfnamefont {H.~P.}\ \bibnamefont {Pfeiffer}}, \ and\
  \bibinfo {author} {\bibfnamefont {T.}~\bibnamefont {Chu}},\ }\href {\doibase
  10.1103/PhysRevD.78.084017} {\bibfield  {journal} {\bibinfo  {journal}
  {Phys.\ Rev.\ D}\ }\textbf {\bibinfo {volume} {78}},\ \bibinfo {pages}
  {084017} (\bibinfo {year} {2008})}\BibitemShut {NoStop}%
\bibitem [{\citenamefont {Cook}\ and\ \citenamefont
  {Pfeiffer}(2004)}]{Cook2004}%
  \BibitemOpen
  \bibfield  {author} {\bibinfo {author} {\bibfnamefont {G.~B.}\ \bibnamefont
  {Cook}}\ and\ \bibinfo {author} {\bibfnamefont {H.~P.}\ \bibnamefont
  {Pfeiffer}},\ }\href {\doibase 10.1103/PhysRevD.70.104016} {\bibfield
  {journal} {\bibinfo  {journal} {Phys.\ Rev.\ D}\ }\textbf {\bibinfo {volume}
  {70}},\ \bibinfo {pages} {104016} (\bibinfo {year} {2004})}\BibitemShut
  {NoStop}%
\bibitem [{\citenamefont {Lindblom}\ \emph {et~al.}(2006)\citenamefont
  {Lindblom}, \citenamefont {Scheel}, \citenamefont {Kidder}, \citenamefont
  {Owen},\ and\ \citenamefont {Rinne}}]{Lindblom:2007}%
  \BibitemOpen
  \bibfield  {author} {\bibinfo {author} {\bibfnamefont {L.}~\bibnamefont
  {Lindblom}}, \bibinfo {author} {\bibfnamefont {M.~A.}\ \bibnamefont
  {Scheel}}, \bibinfo {author} {\bibfnamefont {L.~E.}\ \bibnamefont {Kidder}},
  \bibinfo {author} {\bibfnamefont {R.}~\bibnamefont {Owen}}, \ and\ \bibinfo
  {author} {\bibfnamefont {O.}~\bibnamefont {Rinne}},\ }\href {\doibase
  10.1088/0264-9381/23/16/S09} {\bibfield  {journal} {\bibinfo  {journal}
  {Class.\ Quantum Grav.}\ }\textbf {\bibinfo {volume} {23}},\ \bibinfo {pages}
  {S447} (\bibinfo {year} {2006})},\ \Eprint
  {http://arxiv.org/abs/gr-qc/0512093v3} {arXiv:gr-qc/0512093v3 [gr-qc]}
  \BibitemShut {NoStop}%
\bibitem [{\citenamefont {Friedrich}(1985)}]{Friedrich1985}%
  \BibitemOpen
  \bibfield  {author} {\bibinfo {author} {\bibfnamefont {H.}~\bibnamefont
  {Friedrich}},\ }\href {\doibase 10.1007/BF01217728} {\bibfield  {journal}
  {\bibinfo  {journal} {Commun.\ Math.\ Phys.}\ }\textbf {\bibinfo {volume}
  {100}},\ \bibinfo {pages} {525} (\bibinfo {year} {1985})}\BibitemShut
  {NoStop}%
\bibitem [{\citenamefont {Garfinkle}(2002)}]{Garfinkle2002}%
  \BibitemOpen
  \bibfield  {author} {\bibinfo {author} {\bibfnamefont {D.}~\bibnamefont
  {Garfinkle}},\ }\href@noop {} {\bibfield  {journal} {\bibinfo  {journal}
  {Phys.\ Rev.\ D}\ }\textbf {\bibinfo {volume} {65}},\ \bibinfo {pages}
  {044029} (\bibinfo {year} {2002})}\BibitemShut {NoStop}%
\bibitem [{\citenamefont {{Pretorius}}(2005)}]{Pretorius2005c}%
  \BibitemOpen
  \bibfield  {author} {\bibinfo {author} {\bibfnamefont {F.}~\bibnamefont
  {{Pretorius}}},\ }\href {\doibase 10.1088/0264-9381/22/2/014} {\bibfield
  {journal} {\bibinfo  {journal} {Class.\ Quantum Grav.}\ }\textbf {\bibinfo
  {volume} {22}},\ \bibinfo {pages} {425} (\bibinfo {year} {2005})},\ \Eprint
  {http://arxiv.org/abs/gr-qc/0407110} {gr-qc/0407110} \BibitemShut {NoStop}%
\bibitem [{\citenamefont {{M. A. Scheel, M. Boyle, T. Chu, L. E. Kidder, K. D.
  Matthews and H. P. Pfeiffer}}(2009)}]{Scheel2009}%
  \BibitemOpen
  \bibfield  {author} {\bibinfo {author} {\bibnamefont {{M. A. Scheel, M.
  Boyle, T. Chu, L. E. Kidder, K. D. Matthews and H. P. Pfeiffer}}},\
  }\href@noop {} {\bibfield  {journal} {\bibinfo  {journal} {Phys.\ Rev.\ D}\
  }\textbf {\bibinfo {volume} {79}},\ \bibinfo {pages} {024003} (\bibinfo
  {year} {2009})},\ \Eprint {http://arxiv.org/abs/arXiv:gr-qc/0810.1767}
  {arXiv:gr-qc/0810.1767} \BibitemShut {NoStop}%
\bibitem [{\citenamefont {Lindblom}\ and\ \citenamefont
  {Szil\'agyi}(2009)}]{Lindblom2009c}%
  \BibitemOpen
  \bibfield  {author} {\bibinfo {author} {\bibfnamefont {L.}~\bibnamefont
  {Lindblom}}\ and\ \bibinfo {author} {\bibfnamefont {B.}~\bibnamefont
  {Szil\'agyi}},\ }\href@noop {} {\bibfield  {journal} {\bibinfo  {journal}
  {Phys.\ Rev.\ D}\ }\textbf {\bibinfo {volume} {80}},\ \bibinfo {pages}
  {084019} (\bibinfo {year} {2009})},\ \Eprint
  {http://arxiv.org/abs/arXiv:0904.4873} {arXiv:0904.4873} \BibitemShut
  {NoStop}%
\bibitem [{\citenamefont {Choptuik}\ and\ \citenamefont
  {Pretorius}(2010)}]{Choptuik:2009ww}%
  \BibitemOpen
  \bibfield  {author} {\bibinfo {author} {\bibfnamefont {M.~W.}\ \bibnamefont
  {Choptuik}}\ and\ \bibinfo {author} {\bibfnamefont {F.}~\bibnamefont
  {Pretorius}},\ }\href {\doibase 10.1103/PhysRevLett.104.111101} {\bibfield
  {journal} {\bibinfo  {journal} {Phys.\ Rev.\ Lett.}\ }\textbf {\bibinfo
  {volume} {104}},\ \bibinfo {pages} {111101} (\bibinfo {year} {2010})},\
  \Eprint {http://arxiv.org/abs/0908.1780} {arXiv:0908.1780 [gr-qc]}
  \BibitemShut {NoStop}%
\bibitem [{\citenamefont {{Szil{\'a}gyi}}\ \emph {et~al.}(2009)\citenamefont
  {{Szil{\'a}gyi}}, \citenamefont {{Lindblom}},\ and\ \citenamefont
  {{Scheel}}}]{Szilagyi:2009qz}%
  \BibitemOpen
  \bibfield  {author} {\bibinfo {author} {\bibfnamefont {B.}~\bibnamefont
  {{Szil{\'a}gyi}}}, \bibinfo {author} {\bibfnamefont {L.}~\bibnamefont
  {{Lindblom}}}, \ and\ \bibinfo {author} {\bibfnamefont {M.~A.}\ \bibnamefont
  {{Scheel}}},\ }\href {\doibase 10.1103/PhysRevD.80.124010} {\bibfield
  {journal} {\bibinfo  {journal} {Phys.\ Rev.\ D}\ }\textbf {\bibinfo {volume}
  {80}},\ \bibinfo {eid} {124010} (\bibinfo {year} {2009})},\ \Eprint
  {http://arxiv.org/abs/0909.3557} {arXiv:0909.3557 [gr-qc]} \BibitemShut
  {NoStop}%
\bibitem [{\citenamefont {Cook}\ and\ \citenamefont
  {Scheel}(1997)}]{cook_scheel97}%
  \BibitemOpen
  \bibfield  {author} {\bibinfo {author} {\bibfnamefont {G.~B.}\ \bibnamefont
  {Cook}}\ and\ \bibinfo {author} {\bibfnamefont {M.~A.}\ \bibnamefont
  {Scheel}},\ }\href@noop {} {\bibfield  {journal} {\bibinfo  {journal} {Phys.\
  Rev.\ D}\ }\textbf {\bibinfo {volume} {56}},\ \bibinfo {pages} {4775}
  (\bibinfo {year} {1997})}\BibitemShut {NoStop}%
\bibitem [{\citenamefont {Varma}\ and\ \citenamefont
  {Scheel}(2018)}]{Varma:2018sbhdh}%
  \BibitemOpen
  \bibfield  {author} {\bibinfo {author} {\bibfnamefont {V.}~\bibnamefont
  {Varma}}\ and\ \bibinfo {author} {\bibfnamefont {M.~A.}\ \bibnamefont
  {Scheel}},\ }\href {\doibase 10.1103/PhysRevD.98.084032} {\bibfield
  {journal} {\bibinfo  {journal} {Phys. Rev.}\ }\textbf {\bibinfo {volume}
  {D98}},\ \bibinfo {pages} {084032} (\bibinfo {year} {2018})},\ \Eprint
  {http://arxiv.org/abs/1808.07490} {arXiv:1808.07490 [gr-qc]} \BibitemShut
  {NoStop}%
\bibitem [{\citenamefont {Baumgarte}\ and\ \citenamefont
  {Shapiro}(1999)}]{Baumgarte99}%
  \BibitemOpen
  \bibfield  {author} {\bibinfo {author} {\bibfnamefont {T.~W.}\ \bibnamefont
  {Baumgarte}}\ and\ \bibinfo {author} {\bibfnamefont {S.~L.}\ \bibnamefont
  {Shapiro}},\ }\href@noop {} {\bibfield  {journal} {\bibinfo  {journal}
  {Phys.\ Rev.\ D}\ }\textbf {\bibinfo {volume} {59}},\ \bibinfo {pages}
  {024007} (\bibinfo {year} {1999})},\ \bibinfo {note}
  {gr-qc/9810065}\BibitemShut {NoStop}%
\bibitem [{\citenamefont {Br{\"u}gmann}\ \emph {et~al.}(2008)\citenamefont
  {Br{\"u}gmann}, \citenamefont {Gonz\'{a}lez}, \citenamefont {Hannam},
  \citenamefont {Husa}, \citenamefont {Sperhake},\ and\ \citenamefont
  {Tichy}}]{Bruegmann2006}%
  \BibitemOpen
  \bibfield  {author} {\bibinfo {author} {\bibfnamefont {B.}~\bibnamefont
  {Br{\"u}gmann}}, \bibinfo {author} {\bibfnamefont {J.~A.}\ \bibnamefont
  {Gonz\'{a}lez}}, \bibinfo {author} {\bibfnamefont {M.}~\bibnamefont
  {Hannam}}, \bibinfo {author} {\bibfnamefont {S.}~\bibnamefont {Husa}},
  \bibinfo {author} {\bibfnamefont {U.}~\bibnamefont {Sperhake}}, \ and\
  \bibinfo {author} {\bibfnamefont {W.}~\bibnamefont {Tichy}},\ }\href
  {\doibase 10.1103/PhysRevD.77.024027} {\bibfield  {journal} {\bibinfo
  {journal} {Phys.\ Rev.\ D}\ }\textbf {\bibinfo {volume} {77}},\ \bibinfo
  {eid} {024027} (\bibinfo {year} {2008})},\ \Eprint
  {http://arxiv.org/abs/gr-qc/0610128} {gr-qc/0610128} \BibitemShut {NoStop}%
\bibitem [{\citenamefont {Zlochower}\ \emph {et~al.}(2005)\citenamefont
  {Zlochower}, \citenamefont {Baker}, \citenamefont {Campanelli},\ and\
  \citenamefont {Lousto}}]{Zlochower:2005bj}%
  \BibitemOpen
  \bibfield  {author} {\bibinfo {author} {\bibfnamefont {Y.}~\bibnamefont
  {Zlochower}}, \bibinfo {author} {\bibfnamefont {J.}~\bibnamefont {Baker}},
  \bibinfo {author} {\bibfnamefont {M.}~\bibnamefont {Campanelli}}, \ and\
  \bibinfo {author} {\bibfnamefont {C.}~\bibnamefont {Lousto}},\ }\href
  {\doibase 10.1103/PhysRevD.72.024021} {\bibfield  {journal} {\bibinfo
  {journal} {Phys.\ Rev.\ D}\ }\textbf {\bibinfo {volume} {72}},\ \bibinfo
  {pages} {024021} (\bibinfo {year} {2005})},\ \Eprint
  {http://arxiv.org/abs/gr-qc/0505055} {arXiv:gr-qc/0505055 [gr-qc]}
  \BibitemShut {NoStop}%
\bibitem [{\citenamefont {Sperhake}(2007)}]{Sperhake2006}%
  \BibitemOpen
  \bibfield  {author} {\bibinfo {author} {\bibfnamefont {U.}~\bibnamefont
  {Sperhake}},\ }\href@noop {} {\bibfield  {journal} {\bibinfo  {journal}
  {Phys.\ Rev.\ D}\ }\textbf {\bibinfo {volume} {76}},\ \bibinfo {pages}
  {104015} (\bibinfo {year} {2007})},\ \Eprint
  {http://arxiv.org/abs/gr-qc/0606079} {gr-qc/0606079} \BibitemShut {NoStop}%
\bibitem [{\citenamefont {Pollney}\ \emph {et~al.}(2011)\citenamefont
  {Pollney}, \citenamefont {Reisswig}, \citenamefont {Schnetter}, \citenamefont
  {Dorband},\ and\ \citenamefont {Diener}}]{Pollney:2009yz}%
  \BibitemOpen
  \bibfield  {author} {\bibinfo {author} {\bibfnamefont {D.}~\bibnamefont
  {Pollney}}, \bibinfo {author} {\bibfnamefont {C.}~\bibnamefont {Reisswig}},
  \bibinfo {author} {\bibfnamefont {E.}~\bibnamefont {Schnetter}}, \bibinfo
  {author} {\bibfnamefont {N.}~\bibnamefont {Dorband}}, \ and\ \bibinfo
  {author} {\bibfnamefont {P.}~\bibnamefont {Diener}},\ }\href {\doibase
  10.1103/PhysRevD.83.044045} {\bibfield  {journal} {\bibinfo  {journal}
  {Phys.\ Rev.\ D}\ }\textbf {\bibinfo {volume} {83}},\ \bibinfo {pages}
  {044045} (\bibinfo {year} {2011})}\BibitemShut {NoStop}%
\bibitem [{\citenamefont {Herrmann}\ \emph {et~al.}(2007)\citenamefont
  {Herrmann}, \citenamefont {Hinder}, \citenamefont {Shoemaker},\ and\
  \citenamefont {Laguna}}]{Herrmann2007b}%
  \BibitemOpen
  \bibfield  {author} {\bibinfo {author} {\bibfnamefont {F.}~\bibnamefont
  {Herrmann}}, \bibinfo {author} {\bibfnamefont {I.}~\bibnamefont {Hinder}},
  \bibinfo {author} {\bibfnamefont {D.}~\bibnamefont {Shoemaker}}, \ and\
  \bibinfo {author} {\bibfnamefont {P.}~\bibnamefont {Laguna}},\ }\href@noop {}
  {\bibfield  {journal} {\bibinfo  {journal} {Class.\ Quantum Grav.}\ }\textbf
  {\bibinfo {volume} {24}},\ \bibinfo {pages} {S33} (\bibinfo {year} {2007})},\
  \Eprint {http://arxiv.org/abs/gr-qc/0601026} {gr-qc/0601026} \BibitemShut
  {NoStop}%
\bibitem [{\citenamefont {Pekowsky}\ \emph {et~al.}(2013)\citenamefont
  {Pekowsky}, \citenamefont {O'Shaughnessy}, \citenamefont {Healy},\ and\
  \citenamefont {Shoemaker}}]{Pekowsky:2013ska}%
  \BibitemOpen
  \bibfield  {author} {\bibinfo {author} {\bibfnamefont {L.}~\bibnamefont
  {Pekowsky}}, \bibinfo {author} {\bibfnamefont {R.}~\bibnamefont
  {O'Shaughnessy}}, \bibinfo {author} {\bibfnamefont {J.}~\bibnamefont
  {Healy}}, \ and\ \bibinfo {author} {\bibfnamefont {D.}~\bibnamefont
  {Shoemaker}},\ }\href@noop {} {\bibfield  {journal} {\bibinfo  {journal}
  {Phys.\ Rev.\ D}\ }\textbf {\bibinfo {volume} {88}},\ \bibinfo {pages}
  {024040} (\bibinfo {year} {2013})},\ \Eprint {http://arxiv.org/abs/1304.3176}
  {arXiv:1304.3176 [gr-qc]} \BibitemShut {NoStop}%
\bibitem [{\citenamefont {Brandt}\ and\ \citenamefont
  {Br{\"u}gmann}(1997)}]{Brandt1997}%
  \BibitemOpen
  \bibfield  {author} {\bibinfo {author} {\bibfnamefont {S.}~\bibnamefont
  {Brandt}}\ and\ \bibinfo {author} {\bibfnamefont {B.}~\bibnamefont
  {Br{\"u}gmann}},\ }\href@noop {} {\bibfield  {journal} {\bibinfo  {journal}
  {Phys.\ Rev.\ Lett.}\ }\textbf {\bibinfo {volume} {78}},\ \bibinfo {pages}
  {3606} (\bibinfo {year} {1997})}\BibitemShut {NoStop}%
\bibitem [{\citenamefont {Shibata}\ and\ \citenamefont
  {Nakamura}(1995)}]{shibata95}%
  \BibitemOpen
  \bibfield  {author} {\bibinfo {author} {\bibfnamefont {M.}~\bibnamefont
  {Shibata}}\ and\ \bibinfo {author} {\bibfnamefont {T.}~\bibnamefont
  {Nakamura}},\ }\href@noop {} {\bibfield  {journal} {\bibinfo  {journal}
  {Phys.\ Rev.\ D}\ }\textbf {\bibinfo {volume} {52}},\ \bibinfo {pages} {5428}
  (\bibinfo {year} {1995})}\BibitemShut {NoStop}%
\bibitem [{\citenamefont {Nakamura}\ \emph {et~al.}(1987)\citenamefont
  {Nakamura}, \citenamefont {Oohara},\ and\ \citenamefont {Kojima}}]{NOK87}%
  \BibitemOpen
  \bibfield  {author} {\bibinfo {author} {\bibfnamefont {T.}~\bibnamefont
  {Nakamura}}, \bibinfo {author} {\bibfnamefont {K.}~\bibnamefont {Oohara}}, \
  and\ \bibinfo {author} {\bibfnamefont {Y.}~\bibnamefont {Kojima}},\ }\href
  {\doibase 10.1143/PTPS.90.1} {\bibfield  {journal} {\bibinfo  {journal}
  {Prog. Theor. Phys. Suppl.}\ }\textbf {\bibinfo {volume} {90}},\ \bibinfo
  {pages} {1} (\bibinfo {year} {1987})}\BibitemShut {NoStop}%
\bibitem [{\citenamefont {Baumgarte}\ and\ \citenamefont
  {Shapiro}(2010)}]{baumgarteShapiroBook}%
  \BibitemOpen
  \bibfield  {author} {\bibinfo {author} {\bibfnamefont {T.~W.}\ \bibnamefont
  {Baumgarte}}\ and\ \bibinfo {author} {\bibfnamefont {S.~L.}\ \bibnamefont
  {Shapiro}},\ }\href {\doibase 10.1080/00107514.2011.586052} {\emph {\bibinfo
  {title} {Numerical Relativity: Solving Einstein's Equations on the
  Computer}}}\ (\bibinfo  {publisher} {Cambridge University Press},\ \bibinfo
  {address} {New York},\ \bibinfo {year} {2010})\BibitemShut {NoStop}%
\bibitem [{\citenamefont {Pfeiffer}(2005)}]{Pfeiffer:2005}%
  \BibitemOpen
  \bibfield  {author} {\bibinfo {author} {\bibfnamefont {H.~P.}\ \bibnamefont
  {Pfeiffer}},\ }\href@noop {} {\bibfield  {journal} {\bibinfo  {journal} {J.\
  Hyperbol.\ Differ.\ Eq.}\ }\textbf {\bibinfo {volume} {2}},\ \bibinfo {pages}
  {497} (\bibinfo {year} {2005})},\ \Eprint
  {http://arxiv.org/abs/gr-qc/0412002} {gr-qc/0412002} \BibitemShut {NoStop}%
\bibitem [{\citenamefont {Cook}(2000)}]{Cook2000}%
  \BibitemOpen
  \bibfield  {author} {\bibinfo {author} {\bibfnamefont {G.}~\bibnamefont
  {Cook}},\ }\href {http://www.livingreviews.org/lrr-2000-5} {\bibfield
  {journal} {\bibinfo  {journal} {Living Rev.\ Rel.}\ }\textbf {\bibinfo
  {volume} {3}} (\bibinfo {year} {2000})},\ \bibinfo {note} {5}\BibitemShut
  {NoStop}%
\bibitem [{\citenamefont {Lovelace}(2009)}]{Lovelace2009}%
  \BibitemOpen
  \bibfield  {author} {\bibinfo {author} {\bibfnamefont {G.}~\bibnamefont
  {Lovelace}},\ }\bibfield  {booktitle} {\emph {\bibinfo {booktitle}
  {{Numerical relativity data analysis. Proceedings, 2nd Meeting, NRDA 2008,
  Syracuse, USA, August 11-14, 2008}}},\ }\href {\doibase
  10.1088/0264-9381/26/11/114002} {\bibfield  {journal} {\bibinfo  {journal}
  {Class. Quant. Grav.}\ }\textbf {\bibinfo {volume} {26}},\ \bibinfo {pages}
  {114002} (\bibinfo {year} {2009})},\ \Eprint {http://arxiv.org/abs/0812.3132}
  {arXiv:0812.3132 [gr-qc]} \BibitemShut {NoStop}%
\bibitem [{\citenamefont {Pfeiffer}\ \emph {et~al.}(2007)\citenamefont
  {Pfeiffer}, \citenamefont {Brown}, \citenamefont {Kidder}, \citenamefont
  {Lindblom}, \citenamefont {Lovelace},\ and\ \citenamefont
  {Scheel}}]{Pfeiffer-Brown-etal:2007}%
  \BibitemOpen
  \bibfield  {author} {\bibinfo {author} {\bibfnamefont {H.~P.}\ \bibnamefont
  {Pfeiffer}}, \bibinfo {author} {\bibfnamefont {D.~A.}\ \bibnamefont {Brown}},
  \bibinfo {author} {\bibfnamefont {L.~E.}\ \bibnamefont {Kidder}}, \bibinfo
  {author} {\bibfnamefont {L.}~\bibnamefont {Lindblom}}, \bibinfo {author}
  {\bibfnamefont {G.}~\bibnamefont {Lovelace}}, \ and\ \bibinfo {author}
  {\bibfnamefont {M.~A.}\ \bibnamefont {Scheel}},\ }\href@noop {} {\bibfield
  {journal} {\bibinfo  {journal} {Class.\ Quantum Grav.}\ }\textbf {\bibinfo
  {volume} {24}},\ \bibinfo {pages} {S59} (\bibinfo {year} {2007})},\ \Eprint
  {http://arxiv.org/abs/gr-qc/0702106} {gr-qc/0702106} \BibitemShut {NoStop}%
\bibitem [{\citenamefont {Cook}(2002)}]{Cook2002}%
  \BibitemOpen
  \bibfield  {author} {\bibinfo {author} {\bibfnamefont {G.~B.}\ \bibnamefont
  {Cook}},\ }\href {\doibase 10.1103/PhysRevD.65.084003} {\bibfield  {journal}
  {\bibinfo  {journal} {Phys.\ Rev.\ D}\ }\textbf {\bibinfo {volume} {65}},\
  \bibinfo {pages} {084003} (\bibinfo {year} {2002})}\BibitemShut {NoStop}%
\bibitem [{\citenamefont {Ashtekar}\ and\ \citenamefont
  {Krishnan}(2004)}]{Ashtekar-Krishnan:2004}%
  \BibitemOpen
  \bibfield  {author} {\bibinfo {author} {\bibfnamefont {A.}~\bibnamefont
  {Ashtekar}}\ and\ \bibinfo {author} {\bibfnamefont {B.}~\bibnamefont
  {Krishnan}},\ }\href {http://www.livingreviews.org/lrr-2004-10} {\bibfield
  {journal} {\bibinfo  {journal} {Living Rev.\ Rel.}\ }\textbf {\bibinfo
  {volume} {7}} (\bibinfo {year} {2004})},\ \bibinfo {note} {10}\BibitemShut
  {NoStop}%
\bibitem [{\citenamefont {Dreyer}\ \emph {et~al.}(2003)\citenamefont {Dreyer},
  \citenamefont {Krishnan}, \citenamefont {Shoemaker},\ and\ \citenamefont
  {Schnetter}}]{Dreyer2003}%
  \BibitemOpen
  \bibfield  {author} {\bibinfo {author} {\bibfnamefont {O.}~\bibnamefont
  {Dreyer}}, \bibinfo {author} {\bibfnamefont {B.}~\bibnamefont {Krishnan}},
  \bibinfo {author} {\bibfnamefont {D.}~\bibnamefont {Shoemaker}}, \ and\
  \bibinfo {author} {\bibfnamefont {E.}~\bibnamefont {Schnetter}},\ }\href
  {\doibase 10.1103/PhysRevD.67.024018} {\bibfield  {journal} {\bibinfo
  {journal} {Phys.\ Rev.\ D}\ }\textbf {\bibinfo {volume} {67}},\ \bibinfo
  {pages} {024018} (\bibinfo {year} {2003})}\BibitemShut {NoStop}%
\bibitem [{\citenamefont {{Ossokine}}\ \emph {et~al.}(2015)\citenamefont
  {{Ossokine}}, \citenamefont {{Foucart}}, \citenamefont {{Pfeiffer}},
  \citenamefont {{Boyle}},\ and\ \citenamefont
  {{Szil{\'a}gyi}}}]{Ossokine:2015yla}%
  \BibitemOpen
  \bibfield  {author} {\bibinfo {author} {\bibfnamefont {S.}~\bibnamefont
  {{Ossokine}}}, \bibinfo {author} {\bibfnamefont {F.}~\bibnamefont
  {{Foucart}}}, \bibinfo {author} {\bibfnamefont {H.~P.}\ \bibnamefont
  {{Pfeiffer}}}, \bibinfo {author} {\bibfnamefont {M.}~\bibnamefont {{Boyle}}},
  \ and\ \bibinfo {author} {\bibfnamefont {B.}~\bibnamefont {{Szil{\'a}gyi}}},\
  }\href {\doibase 10.1088/0264-9381/32/24/245010} {\bibfield  {journal}
  {\bibinfo  {journal} {Class.\ Quantum Grav.}\ }\textbf {\bibinfo {volume}
  {32}},\ \bibinfo {eid} {245010} (\bibinfo {year} {2015})},\ \Eprint
  {http://arxiv.org/abs/1506.01689} {arXiv:1506.01689 [gr-qc]} \BibitemShut
  {NoStop}%
\bibitem [{\citenamefont {Buchman}\ \emph {et~al.}(2012)\citenamefont
  {Buchman}, \citenamefont {Pfeiffer}, \citenamefont {Scheel},\ and\
  \citenamefont {Szil{\' a}gyi}}]{Buchman:2012dw}%
  \BibitemOpen
  \bibfield  {author} {\bibinfo {author} {\bibfnamefont {L.~T.}\ \bibnamefont
  {Buchman}}, \bibinfo {author} {\bibfnamefont {H.~P.}\ \bibnamefont
  {Pfeiffer}}, \bibinfo {author} {\bibfnamefont {M.~A.}\ \bibnamefont
  {Scheel}}, \ and\ \bibinfo {author} {\bibfnamefont {B.}~\bibnamefont {Szil{\'
  a}gyi}},\ }\href@noop {} {\bibfield  {journal} {\bibinfo  {journal} {Phys.\
  Rev.\ D}\ }\textbf {\bibinfo {volume} {86}},\ \bibinfo {pages} {084033}
  (\bibinfo {year} {2012})},\ \Eprint {http://arxiv.org/abs/1206.3015}
  {arXiv:1206.3015 [gr-qc]} \BibitemShut {NoStop}%
\bibitem [{\citenamefont {DeDonder}(1921)}]{deDonder1921}%
  \BibitemOpen
  \bibfield  {author} {\bibinfo {author} {\bibfnamefont {T.}~\bibnamefont
  {DeDonder}},\ }\href@noop {} {\emph {\bibinfo {title} {La Gravifique
  {E}insteinienne}}}\ (\bibinfo  {publisher} {Gunthier-Villars},\ \bibinfo
  {address} {Paris},\ \bibinfo {year} {1921})\BibitemShut {NoStop}%
\bibitem [{\citenamefont {Lanczos}(1922)}]{Lanczos22}%
  \BibitemOpen
  \bibfield  {author} {\bibinfo {author} {\bibfnamefont {C.}~\bibnamefont
  {Lanczos}},\ }\href@noop {} {\bibfield  {journal} {\bibinfo  {journal} {Phys.
  Z.}\ }\textbf {\bibinfo {volume} {23}},\ \bibinfo {pages} {537} (\bibinfo
  {year} {1922})}\BibitemShut {NoStop}%
\bibitem [{\citenamefont {Four\`es-Bruhat}(1952)}]{Choquet1952}%
  \BibitemOpen
  \bibfield  {author} {\bibinfo {author} {\bibfnamefont {Y.}~\bibnamefont
  {Four\`es-Bruhat}},\ }\href@noop {} {\bibfield  {journal} {\bibinfo
  {journal} {Acta Math.}\ }\textbf {\bibinfo {volume} {88}},\ \bibinfo {pages}
  {141} (\bibinfo {year} {1952})}\BibitemShut {NoStop}%
\bibitem [{\citenamefont {Fischer}\ and\ \citenamefont
  {Marsden}(1972)}]{fischer_marsden72}%
  \BibitemOpen
  \bibfield  {author} {\bibinfo {author} {\bibfnamefont {A.~E.}\ \bibnamefont
  {Fischer}}\ and\ \bibinfo {author} {\bibfnamefont {J.~E.}\ \bibnamefont
  {Marsden}},\ }\href@noop {} {\bibfield  {journal} {\bibinfo  {journal}
  {Commun.\ Math.\ Phys.}\ }\textbf {\bibinfo {volume} {28}},\ \bibinfo {pages}
  {1} (\bibinfo {year} {1972})}\BibitemShut {NoStop}%
\bibitem [{\citenamefont {Buonanno}\ \emph {et~al.}(2011)\citenamefont
  {Buonanno}, \citenamefont {Kidder}, \citenamefont {Mrou\'{e}}, \citenamefont
  {Pfeiffer},\ and\ \citenamefont {Taracchini}}]{Buonanno:2010yk}%
  \BibitemOpen
  \bibfield  {author} {\bibinfo {author} {\bibfnamefont {A.}~\bibnamefont
  {Buonanno}}, \bibinfo {author} {\bibfnamefont {L.~E.}\ \bibnamefont
  {Kidder}}, \bibinfo {author} {\bibfnamefont {A.~H.}\ \bibnamefont
  {Mrou\'{e}}}, \bibinfo {author} {\bibfnamefont {H.~P.}\ \bibnamefont
  {Pfeiffer}}, \ and\ \bibinfo {author} {\bibfnamefont {A.}~\bibnamefont
  {Taracchini}},\ }\href {\doibase 10.1103/PhysRevD.83.104034} {\bibfield
  {journal} {\bibinfo  {journal} {Phys.\ Rev.\ D}\ }\textbf {\bibinfo {volume}
  {83}},\ \bibinfo {pages} {104034} (\bibinfo {year} {2011})},\ \Eprint
  {http://arxiv.org/abs/1012.1549} {arXiv:1012.1549 [gr-qc]} \BibitemShut
  {NoStop}%
\bibitem [{\citenamefont {Pfeiffer}\ \emph {et~al.}(2003)\citenamefont
  {Pfeiffer}, \citenamefont {Kidder}, \citenamefont {Scheel},\ and\
  \citenamefont {Teukolsky}}]{Pfeiffer2003}%
  \BibitemOpen
  \bibfield  {author} {\bibinfo {author} {\bibfnamefont {H.~P.}\ \bibnamefont
  {Pfeiffer}}, \bibinfo {author} {\bibfnamefont {L.~E.}\ \bibnamefont
  {Kidder}}, \bibinfo {author} {\bibfnamefont {M.~A.}\ \bibnamefont {Scheel}},
  \ and\ \bibinfo {author} {\bibfnamefont {S.~A.}\ \bibnamefont {Teukolsky}},\
  }\href {\doibase 10.1016/S0010-4655(02)00847-0} {\bibfield  {journal}
  {\bibinfo  {journal} {Comput.\ Phys.\ Commun.}\ }\textbf {\bibinfo {volume}
  {152}},\ \bibinfo {pages} {253} (\bibinfo {year} {2003})},\ \Eprint
  {http://arxiv.org/abs/gr-qc/0202096} {gr-qc/0202096} \BibitemShut {NoStop}%
\bibitem [{\citenamefont {{Szil{\'a}gyi}}(2014)}]{Szilagyi:2014fna}%
  \BibitemOpen
  \bibfield  {author} {\bibinfo {author} {\bibfnamefont {B.}~\bibnamefont
  {{Szil{\'a}gyi}}},\ }\href {\doibase 10.1142/S0218271814300146} {\bibfield
  {journal} {\bibinfo  {journal} {{Int. J. Mod. Phys. D}}\ }\textbf {\bibinfo
  {volume} {23}},\ \bibinfo {eid} {1430014} (\bibinfo {year} {2014})},\ \Eprint
  {http://arxiv.org/abs/1405.3693} {arXiv:1405.3693 [gr-qc]} \BibitemShut
  {NoStop}%
\bibitem [{\citenamefont {Boyle}\ and\ \citenamefont
  {Mrou{\'{e}}}(2009)}]{Boyle-Mroue:2008}%
  \BibitemOpen
  \bibfield  {author} {\bibinfo {author} {\bibfnamefont {M.}~\bibnamefont
  {Boyle}}\ and\ \bibinfo {author} {\bibfnamefont {A.~H.}\ \bibnamefont
  {Mrou{\'{e}}}},\ }\href {\doibase 10.1103/PhysRevD.80.124045} {\bibfield
  {journal} {\bibinfo  {journal} {Phys.\ Rev.\ D}\ }\textbf {\bibinfo {volume}
  {80}},\ \bibinfo {pages} {124045} (\bibinfo {year} {2009})},\ \Eprint
  {http://arxiv.org/abs/0905.3177} {arXiv:0905.3177 [gr-qc]} \BibitemShut
  {NoStop}%
\bibitem [{\citenamefont {Blackman}\ \emph {et~al.}(2017)\citenamefont
  {Blackman}, \citenamefont {Field}, \citenamefont {Scheel}, \citenamefont
  {Galley}, \citenamefont {Hemberger}, \citenamefont {Schmidt},\ and\
  \citenamefont {Smith}}]{Blackman:2017dfb}%
  \BibitemOpen
  \bibfield  {author} {\bibinfo {author} {\bibfnamefont {J.}~\bibnamefont
  {Blackman}}, \bibinfo {author} {\bibfnamefont {S.~E.}\ \bibnamefont {Field}},
  \bibinfo {author} {\bibfnamefont {M.~A.}\ \bibnamefont {Scheel}}, \bibinfo
  {author} {\bibfnamefont {C.~R.}\ \bibnamefont {Galley}}, \bibinfo {author}
  {\bibfnamefont {D.~A.}\ \bibnamefont {Hemberger}}, \bibinfo {author}
  {\bibfnamefont {P.}~\bibnamefont {Schmidt}}, \ and\ \bibinfo {author}
  {\bibfnamefont {R.}~\bibnamefont {Smith}},\ }\href {\doibase
  10.1103/PhysRevD.95.104023} {\bibfield  {journal} {\bibinfo  {journal} {Phys.
  Rev.}\ }\textbf {\bibinfo {volume} {D95}},\ \bibinfo {pages} {104023}
  (\bibinfo {year} {2017})},\ \Eprint {http://arxiv.org/abs/1701.00550}
  {arXiv:1701.00550 [gr-qc]} \BibitemShut {NoStop}%
\bibitem [{\citenamefont {Owen}(2009)}]{Owen2009}%
  \BibitemOpen
  \bibfield  {author} {\bibinfo {author} {\bibfnamefont {R.}~\bibnamefont
  {Owen}},\ }\href {\doibase 10.1103/PhysRevD.80.084012} {\bibfield  {journal}
  {\bibinfo  {journal} {Phys.\ Rev.\ D}\ }\textbf {\bibinfo {volume} {80}},\
  \bibinfo {pages} {084012} (\bibinfo {year} {2009})}\BibitemShut {NoStop}%
\bibitem [{\citenamefont {Owen}\ \emph {et~al.}(2011)\citenamefont {Owen},
  \citenamefont {Brink}, \citenamefont {Chen}, \citenamefont {Kaplan},
  \citenamefont {Lovelace}, \citenamefont {Matthews}, \citenamefont {Nichols},
  \citenamefont {Scheel}, \citenamefont {Zhang}, \citenamefont {Zimmerman},\
  and\ \citenamefont {Thorne}}]{OwenEtAl:2011}%
  \BibitemOpen
  \bibfield  {author} {\bibinfo {author} {\bibfnamefont {R.}~\bibnamefont
  {Owen}}, \bibinfo {author} {\bibfnamefont {J.}~\bibnamefont {Brink}},
  \bibinfo {author} {\bibfnamefont {Y.}~\bibnamefont {Chen}}, \bibinfo {author}
  {\bibfnamefont {J.~D.}\ \bibnamefont {Kaplan}}, \bibinfo {author}
  {\bibfnamefont {G.}~\bibnamefont {Lovelace}}, \bibinfo {author}
  {\bibfnamefont {K.~D.}\ \bibnamefont {Matthews}}, \bibinfo {author}
  {\bibfnamefont {D.~A.}\ \bibnamefont {Nichols}}, \bibinfo {author}
  {\bibfnamefont {M.~A.}\ \bibnamefont {Scheel}}, \bibinfo {author}
  {\bibfnamefont {F.}~\bibnamefont {Zhang}}, \bibinfo {author} {\bibfnamefont
  {A.}~\bibnamefont {Zimmerman}}, \ and\ \bibinfo {author} {\bibfnamefont
  {K.~S.}\ \bibnamefont {Thorne}},\ }\href@noop {} {\bibfield  {journal}
  {\bibinfo  {journal} {Phys.\ Rev.\ Lett.}\ }\textbf {\bibinfo {volume}
  {106}},\ \bibinfo {pages} {151101} (\bibinfo {year} {2011})}\BibitemShut
  {NoStop}%
\end{thebibliography}%

\end{document}